\documentclass[12pt]{article}

\usepackage[english]{babel}

\usepackage[utf8]{inputenc}
\usepackage[T1]{fontenc}

\usepackage{afterpage}
\makeatletter
\@fpsep\textheight
\makeatother
\usepackage{setspace}
\usepackage[margin=1in]{geometry}

\usepackage{bm}
\usepackage{amsmath, amssymb}
\usepackage{amsfonts, amsthm}
\usepackage{thmtools}
\usepackage{mathtools}
\usepackage{array}
\allowdisplaybreaks

\usepackage{algorithm, algpseudocode}

\usepackage{graphicx}
\usepackage{epstopdf}
\usepackage[font=sc]{caption}
\usepackage[labelformat=simple]{subcaption}

\usepackage{pgfplots}
\pgfplotsset{compat=1.10}
\usepgfplotslibrary{fillbetween}
\usepackage{tikz}
\newenvironment{fignote}[1][\linewidth]{%
    \begin{center}
    \begin{minipage}{#1}\footnotesize
}{%
    \end{minipage}
    \end{center}
}

\usepackage{threeparttable}
\usepackage{longtable}
\usepackage{multirow}
\usepackage{tabularx}
\usepackage{booktabs}
\usepackage{rotating}
\usepackage{pdflscape}
\usepackage{float}

\usepackage[colorlinks=true, citecolor=blue, urlcolor=blue, linkcolor=blue]{
    hyperref
}

\usepackage[nameinlink]{cleveref}
\usepackage{natbib}

\usepackage[dvipsnames]{xcolor}
\usepackage{enumitem}
\usepackage{authblk} 
\usepackage[title]{appendix}
\usepackage{xr}

\theoremstyle{definition}
\newtheorem{assumption}{Assumption}
\newtheorem{definition}{Definition}
\newtheorem{proposition}{Proposition}
\newtheorem{theorem}{Theorem}

\newtheorem{lemma}{Lemma}
\newtheorem{example}{Example}
\AtBeginEnvironment{example}{%
  \pushQED{\qed}%
}
\AtEndEnvironment{example}{\popQED\endexample}
\newtheoremstyle{boldremark}
    {\dimexpr\topsep/2\relax} 
    {\dimexpr\topsep/2\relax} 
    {}          
    {}          
    {\bfseries} 
    {.}         
    {.5em}      
    {}          

\theoremstyle{boldremark}
\newtheorem{remark}{Remark}

\renewcommand\thmcontinues[1]{Continued}

\newif\ifdraft
\drafttrue        
 \draftfalse      

\title{\sc Beyond Validity: SVAR Identification Through the Proxy Zoo\thanks{Contacts: Jiaming Huang (\href{mailto:jiaminghuang@ust.hk}{jiaminghuang@ust.hk}) and Luca Neri (\href{mailto:luca.neri@uclouvain.be}{luca.neri@uclouvain.be}).
We would like to thank Luca Fanelli, Byoungchan Lee,  and seminar participants at CUHK Shenzhen, 2025 Annual Meeting of the Greater Bay Econometrics Study Group, and CEP Brown Bag at HKUST.}}
\author[1]{Jiaming Huang}
\author[2]{Luca Neri}
\affil[1]{The Hong Kong University of Science and Technology}
\affil[2]{UCLouvain}
\date{January 16, 2026} 

\begin{document}
\maketitle

\begin{abstract}
    \setstretch{1}\noindent
    This paper develops a framework for robust identification in SVARs when researchers face a \emph{zoo} of proxy variables.
    Instead of imposing exact exogeneity, we introduce \emph{generalized ranking restrictions} (GRR) that bound the relative correlation of each proxy with the target and non-target shocks through a continuous proxy-quality parameter.
    Combining GRR with standard sign and narrative restrictions, we characterize identified sets for structural impulse responses and show how to partially identify the proxy-quality parameter using the joint information contained in the proxy zoo. We further develop sensitivity and diagnostic tools that allow researchers to assess transparently how empirical conclusions depend on proxy exogeneity assumptions and the composition of the proxy zoo.
    A simulation study shows that proxies constructed from sign restrictions can induce biased proxy-SVAR estimates, while our approach delivers informative and robust identified sets.
    An application to U.S.\ monetary policy illustrates the empirical relevance and computational tractability of the framework.


\end{abstract}

\doublespacing
\newpage
\section{Introduction}

Identification through external instruments---or \emph{proxies}---is widely used in structural vector autoregressions (SVARs). By exploiting variables that are correlated with a specific structural shock while remaining orthogonal to all others, the proxy-SVAR approach delivers point-identified impulse responses while maintaining a flexible reduced-form specification \citep{mertens2013dynamic,stock2018identification}. This approach has been adopted across a wide range of applications, including monetary policy \citep{gertler2015monetary}, fiscal policy \citep{mertens2013dynamic}, oil shocks \citep{kanzig2021macroeconomic}, uncertainty shocks \citep{angelini2019exogenous}, and financial shocks \citep{ottonello2025financial}, among many others.

\medskip

At the same time, this empirical success has brought into focus a fundamental tension.
The orthogonality requirement is often difficult to justify and impossible to test directly.
This has led to a proliferation of proxies intended to identify the same structural shock---a
\emph{proxy zoo}---constructed using different data sources and methodologies, and to a central debate over proxy exogeneity.
Recent work documents non-negligible correlations between widely used proxies and non-target shocks, challenging exact exogeneity as a maintained assumption.
For example, \citet{schlaak2023monetary} raise concerns about the exogeneity of high-frequency monetary policy surprises, while the narrative shocks of \citet{romer2004new} have long been argued to reflect oil-related disturbances \citep[e.g.,][]{barnichon2025innovations}.

\medskip

Instead of resolving disputes over proxy exogeneity, we develop a framework for robust identification in SVARs with a proxy zoo.\footnote{Although we develop the framework in the context of SVARs, the approach extends naturally to local projections with external instruments following \cite{plagborg-moller2021local}.}
Central to identification is a set of \textit{generalized ranking restrictions} (GRR) on the relative correlation of each proxy with the target and non-target shocks, governed by a continuous proxy-quality parameter.
We show three main results.
First, we characterize the identified set for structural impulse responses under GRR combined with standard sign and narrative restrictions.
Second, we show how to partially identify the proxy-quality parameter using the joint information contained in the proxy zoo and these additional identifying restrictions.
Third, we develop a suite of sensitivity and diagnostic tools in the spirit of \citet{manski2003partial} that allow researchers to assess transparently how substantive conclusions depend on assumptions about proxy exogeneity.

The identifying power of the GRR is shaped by two distinct forces: (i) the proxy-quality parameter and (ii) proxy complementarity.
The proxy-quality formulation delivers transparency and robustness. It requires researchers to state explicitly the degree of exogeneity violations they are willing to tolerate, with the conventional valid-IV assumption emerging only as a limiting case with zero contamination. The resulting identified sets contain the true impulse responses whenever the true proxy quality exceeds the imposed level.
Complementarity arises when proxies carry distinct, falsifiable information \citep[cf.][]{masten2021salvaging}, restricting different directions in the space of admissible structural representations. Strikingly, when proxies are sufficiently complementary, point identification can be recovered even when all of them are contaminated.\footnote{Because informative identification is possible without valid instruments, the framework encourages the construction of new proxies under weaker conditions than classical exogeneity, broadening the scope of external information that can be brought to bear on structural identification.}

Crucially, the quality parameter governing the GRR need not be calibrated ex ante. Instead, we exploit two sources of falsifiable information in the proxy zoo to derive an application-specific upper bound on proxy quality.
First, once auxiliary economic restrictions, such as sign or narrative constraints, are imposed, the data themselves restrict the range of contamination levels compatible with those restrictions.
Second, when proxies contain conflicting information, each proxy rules out structural representations that others would admit, so that joint feasibility disciplines admissible contamination levels without parametric modeling of endogeneity \citep[see e.g.,][]{nguyen2025bayesian}.

\medskip
Having established an upper bound on proxy quality, we recommend reporting two diagnostic measures.
First, the breakdown frontier, which records the minimum level of proxy quality required to sustain empirical claims of interest.
Second, proxy informativeness, which quantifies the contribution of each proxy to tightening the identified set and detects redundancy among proxies.
These diagnostics complement standard sensitivity analysis by clarifying how empirical conclusions depend on the composition of the proxy zoo, thereby connecting proxy-SVAR identification to recent advances in sensitivity analysis and research transparency \citep{vanderweele2017sensitivity,andrews2020transparency,masten2020inference}. In our monetary policy application, these diagnostics reveal substantial heterogeneity in the identifying content of commonly used proxies, with a small subset driving most of the tightening of the identified set and several others contributing little additional information once the proxy zoo is considered jointly.

\medskip
A simulation study based on a medium-scale dynamic stochastic general equilibrium (DSGE) model \citep{smets2007shocks} further demonstrates that treating contaminated proxies as valid instruments---such as proxies constructed from sign-restriction procedures \citep[e.g.,][]{baumeister2019structural,jarocinski2020deconstructing}---can produce substantial bias while preserving the expected sign of responses, making contamination difficult to detect from ``puzzles'' alone.\footnote{The issue does not stem from sign restrictions as an identification device, but from common procedures that map sign-restricted SVARs into a single proxy series for external use. These constructions need not preserve orthogonality with non-target shocks.} In contrast, our method recovers the true impulse responses within robust identified sets.
Empirically, we illustrate the method using a rich set of U.S.\ monetary policy proxies drawn from narrative, high-frequency based, and model-based approaches. We show that conventional proxy-SVAR estimates are sensitive to the choice of proxy, while our set-identified approach delivers informative and transparent bounds even without assuming any proxy to be exogenous. In this application, the data imply a finite upper bound on the proxy-quality parameter, indicating that the joint restrictions imposed by the proxy zoo  are not compatible with exact exogeneity for all proxies simultaneously.

\bigskip
\noindent\emph{Literature.} Our framework contributes to several strands of the literature.
First, it nests popular identification schemes in SVARs as special cases. By varying the proxy-quality parameter, our GRR reduces to the least restrictive external-variable constraints \citep{ludvigson2021uncertainty}, to the more informative ranking-based identification \citep{braun2023identification}, and to the classical proxy-SVAR framework based on exact exogeneity \citep{mertens2013dynamic}. Our auxiliary restrictions encompass both conventional sign restrictions on impulse responses \citep{uhlig2005what} and narrative-based restrictions \citep{antolin-diaz2018narrative}, making explicit how empirical conclusions depend on identifying assumptions.

Second, the paper relates to recent work addressing proxy contamination while preserving identification. Prominent approaches exploit volatility changes \citep{schlaak2023monetary,angelini2025invalid}, higher-order moment restrictions \citep{keweloh2025estimating}, combinations thereof \citep{carriero2024blended}, or innovations-powered inference \citep{barnichon2025innovations}. Our framework differs by treating all proxies as potentially contaminated and exploiting their joint restrictions for identification of a single shock, rather than relying on auxiliary statistical structure.

Third, our partial identification of proxy quality connects to the literature on overidentification tests, dating back to \citet{sargan1958estimation}, and their recent applications to assess proxy exogeneity \citep{schlaak2023monetary,bruns2024testing,angelini2025test,angelini2025invalid}.
In contrast, we exploit the direction of endogeneity bias---revealed by overidentifying restrictions---to learn about the parameter of interest \cite[cf.][]{masten2021salvaging}.

Finally, the framework extends the proxy-SVAR approach to sensitivity analysis \citep{manski2003partial,andrews2017measuring,masten2020inference}, and relates to recent advances on ``plausibly exogenous'' instruments \citep{conley2012plausibly} and regression sensitivity analysis \citep{kiviet2020testing,cinelli2020making}.

\bigskip
\noindent
\emph{Outline.} The remainder of the paper is organized as follows. Section~\ref{sec:motivation} provides the  motivating descriptive evidence that illustrates potential contamination of popular monetary policy proxies. Section~\ref{sec:framework} presents the general framework for set-identified proxy-SVAR with the generalized ranking restrictions. The proxy quality parameter is partially identified in Section~\ref{sec:bounds_tau}, based on which a set of diagnostic tools for sensitivity analysis is discussed in Section~\ref{sec:sensitivity}.
Section~\ref{sec:simulation} presents the results of a simulation study and Section~\ref{sec:application} revisits the monetary policy application.
Section~\ref{sec:conclusion} concludes.

\section{Motivating descriptive evidence}\label{sec:motivation}

We illustrate the empirical challenges of conventional proxy-SVAR identification using the case of U.S.\ monetary policy shocks.
Rather than assessing the validity of individual proxies, this section documents that exact exogeneity imposed jointly across a proxy zoo is a strong maintained assumption. We show that monetary policy proxies both deliver heterogeneous impulse responses and exhibit nontrivial correlations with proxies for other structural shocks.

We estimate a standard seven-variable monthly VAR with industrial production (INDPRO), CPI inflation (CPIAUCSL), the unemployment rate (UNRATE), the CRB commodity price index (CRBPI), the one-year Treasury rate (GS1), financial market prices (S\&P500), and the excess bond premium (EBP), using data from 1973m1 to 2019m12.\footnote{The specification includes 12 lags and a constant.} The specification follows \citet{miranda-agrippino2021transmission} and is widely used in empirical analyses of U.S.\ monetary policy.

We consider eight external instruments that are among the most widely used in the literature:
the narrative series of \citet{romer2004new} (RR);
the target-factor proxy of \citet{gurkaynak2005actions} (GSST);\footnote{Constructed from the authors' replication of \citet{gurkaynak2005actions}.}
the high-frequency innovation of \citet{gertler2015monetary} (GK);
the series of \citet{nakamura2018highfrequency} (NS);\footnote{Obtained from the supplementary materials of \citet{brennan2025monetary}.}
the unified shock measure of \citet{bu2021unified} (BRW);\footnote{We use the cumulative-sum proxy following \citet{bu2021unified}.}
the proxy of \citet{miranda-agrippino2021transmission} (MR);
the sign-restriction-based measure of \citet{jarocinski2020deconstructing} (JK);
and the high-frequency surprise of \citet{bauer2023reassessment} (BS).
These proxies are constructed under different principles---narrative analysis, high-frequency surprises, factor decomposition---but are all intended to identify the same underlying monetary policy shock, even though they may capture different aspects of policy actions and information releases.

\begin{figure}[ht]
    \centering
    \includegraphics[width=\linewidth]{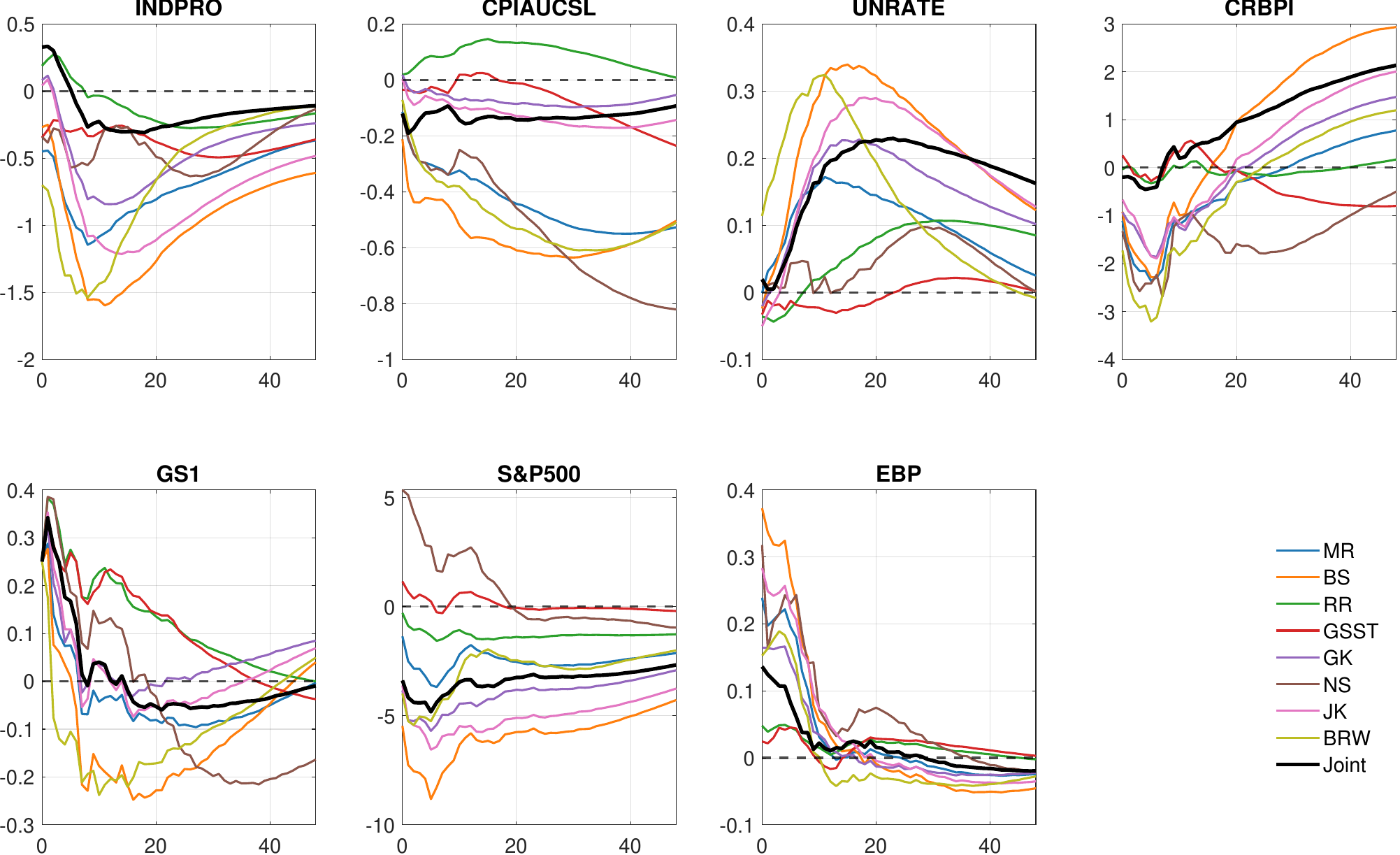}
    \caption{Proxy-SVAR, Point-identified IRFs to a monetary policy shock}

    \begin{fignote}[\linewidth]
        \emph{Notes:} Impulse responses to a contractionary monetary policy shock that increases GS1 by 25bp. The IRFs are identified using alternative external instruments listed in the legend. Each colored line corresponds to a proxy-specific point-identified SVAR. The black solid line reports the impulse response obtained when all proxies are imposed jointly as identifying restrictions.
    \end{fignote}
    \label{fig:proxySVAR_joint}
\end{figure}

\autoref{fig:proxySVAR_joint} reports the impulse responses obtained from point-identified proxy-SVARs when each proxy is used individually. The resulting responses differ substantially across instruments, both in magnitude and in dynamic shape. In some cases, the implied policy rate path displays patterns that are not typically associated with contractionary monetary policy, such as a negative response of short-term interest rates at medium horizons. Such heterogeneity does not permit conclusions about the validity or invalidity of any particular proxy; rather, it highlights the sensitivity of point-identified inference to the choice of instrument.

A key implication of the conventional approach is that, if all proxies satisfied the identifying assumptions and were sufficiently informative, they would deliver identical impulse responses. Substantial discrepancies across instruments therefore indicate that inference based on a single proxy may be fragile in practice. As an illustration, two influential proxies, MR and BS, disagree on the sign of the identified structural shock in 43.75\% of the months between 1979 and 2019; see \autoref{fig:EMP-MRBS-Sign-Disagreement} in the appendix. This disagreement is not diagnostic of invalidity, but it underscores the fragility of inference based on any single proxy and motivates identification strategies that remain informative when proxies provide conflicting signals.

In principle, exogeneity restrictions can be assessed in overidentified systems using tests of overidentifying restrictions \citep{sargan1958estimation,hansen1982large}. However, the interpretation of such tests relies on maintaining that at least one proxy is exactly exogenous \citep[cf.][]{kiviet2020testing}, which becomes a strong maintained assumption when all available proxies may be subject to some degree of contamination.

As complementary descriptive evidence, we examine pairwise correlations between the monetary policy proxy zoo and a broad set of proxies for other structural shocks, including oil supply, financial, fiscal, and technology shocks.\footnote{We consider 19 non-monetary proxies from \cite{kanzig2021macroeconomic}, \cite{baumeister2019structural}, \cite{ottonello2025financial}, \cite{ramey2018government}, \cite{fisher2010using}, \cite{benzeev2017chronicle}, \cite{leeper2012quantitative}, \cite{beaudry2006stock}, \cite{fernald2014quarterly}, \cite{benzeev2015investmentspecific}, \cite{benzeev2018what}, and \cite{francis2014flexible}. Some series are available only at quarterly frequency; for these, we sum the monetary policy instruments within quarters.} Under an ideal benchmark where structural shocks are orthogonal and measurement errors are independent, exogeneity implies zero correlation across proxies for distinct shocks. While non-zero correlations can arise from sampling variation or common information, we treat these patterns as descriptive indicators of potential contamination risks.

\begin{figure}[ht]
    \centering
    \includegraphics[width=\linewidth]{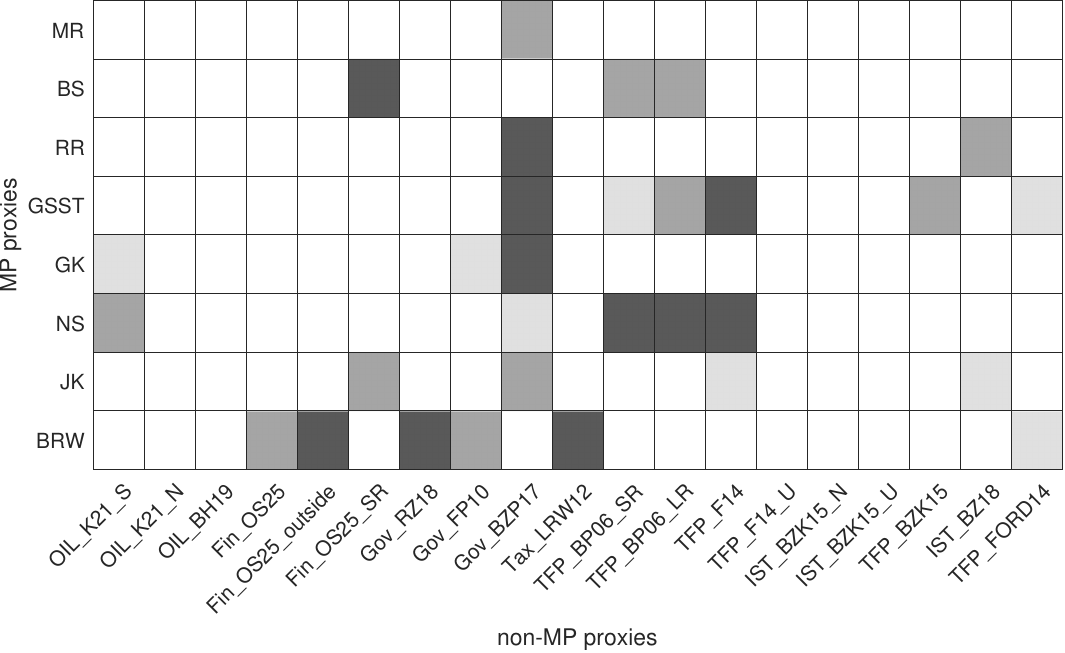}
    \caption{Correlation significance map}

    \begin{fignote}[\linewidth]
        \emph{Notes:} Monetary policy proxies vs.\ non-monetary policy proxies. Light gray: $p\text{\normalfont-val}\leq 0.10$; gray: $p\text{\normalfont-val}\leq 0.05$; dark gray: $p\text{\normalfont-val}\leq 0.01$. We only consider months in which both proxies record a shock. Details of correlation coefficients and non-monetary proxy definitions are provided in \autoref{tab:pairwise_corr} in the appendix.
    \end{fignote}
    \label{fig:correlation_significance_map}
\end{figure}

\autoref{fig:correlation_significance_map} reports the correlation-significance map. Two broad patterns emerge.
First, fiscal and financial shocks appear to be common sources of contamination. Most monetary proxies exhibit strong positive correlations with the government spending news shock of \cite{benzeev2017chronicle} (ranging from 0.15 to 0.43). Correlations are even more prominent with the \cite{ramey2018government} military news shock (ranging from 0.15 to 0.66), except for the RR proxy.
Similarly, most monetary proxies show negative correlations with the sign-restricted financial proxy of \citet{ottonello2025financial}, while some---specifically RR and BRW---also correlate significantly with the high-frequency-based financial proxy.
Second, idiosyncratic contamination patterns are prevalent. For instance, the NS proxy correlates significantly with oil news shocks and various TFP measures, whereas the BRW proxy displays uniquely high correlations with tax shocks (0.56) and defense spending news (0.66).
Taken together, these patterns illustrate that exact exogeneity imposed on individual proxies---or jointly across a proxy zoo---may be delicate in practice, motivating identification strategies that remain informative under limited forms of contamination.

\section{Framework}\label{sec:framework}
Consider a standard $n$-variate SVAR($p$) model, with $p<\infty$:
\begin{equation}
    Y_t = A_1Y_{t-1} + \dots + A_p Y_{t-p} + u_t, \qquad u_t = B\epsilon_t,\label{eq:SVAR}
\end{equation}
where $Y_t$ is the vector of observables, $ u_{t} $ is the vector of reduced-form innovations, $\epsilon_t$ is the vector of structural shocks, and $B$ is the structural matrix capturing the contemporaneous effects of the structural shocks on the variables $Y_t$.
We impose the following standard regularity conditions.

\begin{assumption}[SVAR Assumption \citep{jentsch2022asymptotically}]\label{ass:svar}\hfill
    \begin{enumerate}[label={(\roman*).},ref={\theassumption.(\roman*)}]
        \item The lag order $p$ is known, and $\det(I_n - A_1z - \dots - A_pz^p)$ has all roots outside the unit circle so that the DGP is a stationary and causal VAR model.\label{ass:svar_stationarity}
        \item $(\epsilon_t, \; t \in \mathbb{Z})$ is a white-noise process, that is, $\mathbb{E}(\epsilon_t) = 0$ and $\mathbb{E}(\epsilon_t \epsilon_{t-j}') = 0$ for $j = 0$, with $\mathbb{E}(\epsilon_t \epsilon_t') =I_n$.\label{ass:svar_shock}
        \item $B$ is an invertible ($n\times n$) matrix.\label{ass:svar_bmat}
    \end{enumerate}
\end{assumption}
Assumption \ref{ass:svar} collects standard conditions ensuring that a structural VAR is well-defined. Part \ref{ass:svar_stationarity} imposes stability of the reduced form, which guarantees the existence of a causal VAR representation and well-defined impulse responses.
Part \ref{ass:svar_shock} specifies that structural shocks are orthonormal innovations, which normalizes shock variances to unity and ensures shocks correspond to economically distinct primitive forces.
Finally, Part \ref{ass:svar_bmat} requires $B$ to be invertible. This ensures that the reduced-form innovations have a non-degenerate structural representation and that each structural shock corresponds to a distinct direction in the space of reduced-form errors \citep{stock2018identification}. Identification procedures, including sign restrictions and external instruments, restrict the set of admissible structural matrices.

Without loss of generality, we focus on identifying the dynamic causal effects of the first structural shock $ \epsilon_{1,t} $.
The impulse response of variable $i$ to shock $\epsilon_1$ at horizon $h$ is given by the $(i,1)$-th element of $ C_{h}B $, where
\begin{equation}
    C_0 = I_n, \qquad C_h = \sum_{\ell=1}^{h} C_{h-\ell} A_\ell, \quad h = 1,\ldots,H,
    \label{eq:def_reduced_irf}
\end{equation}
and $A_\ell=0_{n\times n}$ for $\ell>p$; see \citet[][p. 111]{kilian2017structural} for a detailed derivation. Thus, the response of variable $i$ to shock $j$ at horizon $h$ can be written as
\begin{equation}\label{eq:IRF_ijh}
    \theta_{i,j,h} = e_i'C_hBe_{j}
\end{equation}
where $e_i$ denotes the $i$-th elementary basis vector of $\mathbb{R}^n$.
Since each $C_h$ depends only on the autoregressive coefficients, which are consistently estimable by OLS, identification reduces to recovering (a column of) the structural matrix $B$.

\paragraph{Point identification via valid instruments.}
In the presence of a valid instrument $ m_{t} $ for the target shock $\epsilon_{1,t}$, point identification of the first column of $B$ can be achieved via the proxy-SVAR approach \citep[e.g.,][]{mertens2013dynamic}. Specifically, if $ \mathbb{E}[m_{t}\epsilon_{1,t}] \neq 0 $ and $ \mathbb{E}[m_{t}\epsilon_{j,t}] = 0 $ for all $ j \geq 2 $, then
\begin{equation}\label{eq:M_validIV}
    \mathbb{E}[u_{t}m_t] = \mathbb{E}[B\epsilon_tm_t] = B \begin{pmatrix}\mathbb{E}[\epsilon_{1,t}m_t]\\ 0_{n-1}\end{pmatrix} = B_{\bullet 1} \mathbb{E}[\epsilon_{1,t}m_t]~.
\end{equation}
The structural responses to shock $\epsilon_{1,t}$ are thus identified up to a scaling factor by combining the second-moment restrictions with the valid-IV assumption.
However, instrument validity is often questionable in practice. With contamination---$\mathbb{E}[ m_t \epsilon_{j,t}] \neq 0$ for some $j \geq 2$---the standard approach identifies only a linear combination of structural responses:
\begin{equation}\label{eq:cov_contam}
    \mathbb{E}[u_{t}m_t] = B_{\bullet 1} \mathbb{E}[\epsilon_{1,t}m_t] + \dots + B_{\bullet n} \mathbb{E}[\epsilon_{n,t}m_t]~.
\end{equation}
Identification breaks down unless responses to different non-target shocks are proportionally similar, which is unlikely in practice.

\paragraph{Set identification approach.}
Instead of pursuing point identification, we adopt a partial identification framework that characterizes the \textit{set} of structural matrices $B$ compatible with both the statistical properties of the reduced-form VAR and economic restrictions.
The reduced-form VAR model already provides $n(n+1)/2$ second-moment restrictions from the reduced-form covariance matrix:
\begin{equation}
    \Sigma = \mathbb{E}[u_tu_t']= B \mathbb{E}[\epsilon_t\epsilon_t'] B'=BB'~.
\end{equation}
While insufficient to uniquely pin down all $ n^2 $ elements of $ B $, these restrictions provide a useful starting point.
Let $L$ denote the unique lower-triangular Cholesky factor of $\Sigma$ such that $\Sigma = LL'$, and let $\mathcal{O}(n) = \{Q \in \mathbb{R}^{n \times n}: QQ' = I_n\}$ denote the set of orthogonal matrices of dimension $n$. Then any matrix $\tilde{B} = LO$ with $O \in \mathcal{O}(n)$ automatically satisfies the second-moment restrictions:
\begin{equation}
    \Sigma = BB'=LL'=LOO'L'=\tilde{B}\tilde{B}'~.
\end{equation}
Thus, $B$ is identified only up to an orthogonal rotation $O$.
Set identification reduces to characterizing the orthogonal matrices $O$ consistent with additional economic restrictions introduced below.

For future usage, we define the proxy moment vector $M_\ell \coloneqq \mathbb{E}[L^{-1}u_t\, m_{\ell,t}]$, and we collect the VAR coefficients, the covariance elements, and the proxy moments into a single vector:
\begin{equation}
    \phi \coloneqq \big(\text{vec}(A_1,\ldots,A_p)', \text{vech}(\Sigma)', M_1',\ldots,M_k'\big)'~,\label{eq:def_phi}
\end{equation}
where $\text{vec}(\cdot)$ stacks matrix columns and $\text{vech}(\cdot)$ stacks the lower-triangular elements.

\subsection{Generalized ranking restrictions}

We introduce an identification strategy that exploits external proxy variables while relaxing the conventional valid-IV assumption. Let $m_t=(m_{1,t},\ldots,m_{k,t})'$ denote a vector of $k$ external variables intended to contain information about the target shock $\epsilon_{1,t}$. We refer to the collection $(m_{1,t},\ldots,m_{k,t})$ as the \textit{proxy zoo}.

The classical proxy-SVAR framework requires each proxy to be uncorrelated with all non-target shocks, that is, Equation~\eqref{eq:M_validIV} holds. In many applications, however, this assumption is empirically difficult to maintain: variables designed to capture a specific structural shock often respond, at least weakly, to other economic forces. Once validity is relaxed, the relevant question is therefore not whether contamination is present, but whether it is sufficiently limited for the proxy to remain informative about the target shock.

The following example illustrates this issue and motivates our approach.

\begin{example}[Motivating proxy contamination]\label{eg:grr}
    Consider a VAR with three structural shocks—monetary policy ($\epsilon_{\mathrm{mp},t}$), TFP ($\epsilon_{\mathrm{tfp},t}$), and cost-push ($\epsilon_{\mathrm{cp},t}$)—and a generic proxy $m_{\ell,t}$ with population correlations
    \begin{equation}
        \mathrm{corr}(m_{\ell,t},\epsilon_{\mathrm{mp},t}) = 0.3,\quad
        \mathrm{corr}(m_{\ell,t},\epsilon_{\mathrm{tfp},t}) = -0.05,\quad
        \mathrm{corr}(m_{\ell,t},\epsilon_{\mathrm{cp},t}) = 0.10,
    \end{equation}
    where $\epsilon_{\mathrm{mp},t}$ is the target shock. This proxy violates the valid-IV assumption because it is correlated with non-target shocks. Nevertheless, its correlation with the target shock is substantially larger than with any other shock, suggesting that it still contains meaningful directional information.
\end{example}

Example~\ref{eg:grr} makes clear that, once validity is abandoned, identification hinges on the relative strength of a proxy's correlation with the target shock relative to non-target shocks. We formalize the comparison between these correlations through a proxy-specific quality parameter that bounds how strongly a proxy may co-move with non-target shocks relative to the target shock. Importantly, this parameter is not calibrated ex ante: in later sections, we show that it is itself partially identified from the joint restrictions implied by the data and the maintained identifying assumptions.

\begin{assumption}[Proxy properties]\label{ass:proxy}
    For each proxy $m_{\ell,t}$, $\ell=1,\ldots,k$:
    \begin{enumerate}[label={(\roman*).},ref={\theassumption.(\roman*)}]
        \item $\mathbb{E}[m_{\ell,t}]=0$, $\mathbb{E}[m_{\ell,t}^2]<\infty$, and $\mathbb{E}[u_t m_{\ell,t}]$ exists. \label{ass:proxy_finite_moments}
        \item There exists $\tau_{\ell,0}=\min_{j\ge2}\tau_{\ell,j,0}\ge0$ such that for each $j\ge2$ \label{ass:proxy_tau0}
              \begin{equation}
                  \mathrm{corr}(m_{\ell,t},\epsilon_{1,t})
                  =
                  \tau_{\ell,j,0}\,
                  \big|\mathrm{corr}(m_{\ell,t},\epsilon_{j,t})\big| .
                  \label{eq:proxy_contamination}
              \end{equation}
    \end{enumerate}
\end{assumption}

Assumption~\ref{ass:proxy_finite_moments} imposes standard regularity conditions on the proxies, as commonly assumed in the proxy-SVAR literature \citep[e.g.,][]{mertens2013dynamic}.
Assumption~\ref{ass:proxy_tau0} formalizes the intuition in Example~\ref{eg:grr} by parameterizing proxy quality in terms of the relative magnitude of correlations with the target and non-target shocks. Specifically, for a given proxy, the parameter $\tau_{\ell,0}$---defined as the minimum $\tau_{\ell,j,0}$ across non-target shocks---summarizes how large the proxy's correlation with the target shock is relative to its correlations with non-target shocks. In Example~\ref{eg:grr},
$\tau_{\ell,0}=3$, reflecting that the proxy's correlation with the monetary-policy shock exceeds its correlation with each non-target shock by a factor of three, at least. Any finite value of $\tau_{\ell,0}$
permits contamination while indexing how informative the proxy is about the target shock. Assumption~\ref{ass:proxy_tau0} is operationalized as a set of \textit{ranking restrictions} on the rotation matrix
$O.$

\paragraph{Operationalizing the proxy assumption.}
The contamination assumption \eqref{eq:proxy_contamination} is stated in terms of population correlations between proxies and structural shocks, which are unobservable. The next result expresses this as explicit constraints on $O$ and a consistently estimable moment vector.

\begin{proposition}[Generalized ranking restrictions]\label{prop:rank_O}
    Under Assumptions \ref{ass:svar} and \ref{ass:proxy}, the proxy contamination assumption \eqref{eq:proxy_contamination} implies
    \begin{equation}
        O_{\bullet 1}' M_\ell
        \;\ge\;
        \tau_{\ell,0}\,
        \big| O_{\bullet j}' M_\ell \big|,
        \qquad \ell = 1,\ldots,k,\; j\ge 2 .
        \label{eq:rank_O}
    \end{equation}
\end{proposition}

Since $M_\ell$ is consistently estimable, this proposition yields a set of implementable linear inequality constraints on $O$. We refer to the linear inequality constraints in \eqref{eq:rank_O} as the generalized ranking restrictions (GRR).

\begin{example}[Generalized ranking with exogeneity]\label{eg:grr_iv}
    Assumption~\ref{ass:proxy_tau0} nests the standard exogeneity assumption as the limiting case $ \tau_{\ell,0}=\infty $. Under perfect exogeneity, the generalized ranking restrictions \eqref{eq:rank_O} reduce to
    \begin{equation}
        O_{\bullet 1}'M_{\ell} >0,\quad
        O_{\bullet j}'M_{\ell} =0,\quad \ell=1,\ldots,k,;\forall j\ge 2~.
    \end{equation}
    Since the columns of $O$ form an orthonormal basis for $\mathbb{R}^n$, any vector $M_{\ell}$ admits the decomposition $M_{\ell}=\sum_{j=1}^{n}a_{\ell,j}O_{\bullet j}$ for some constants $a_{\ell,1},\dots,a_{\ell,n}$. The orthogonality conditions $O_{\bullet j}'M_{\ell} =0$ for $j \geq 2$ imply $ a_{\ell,j}=0 $ for all $j\ge 2$, yielding $M_{\ell}=a_{\ell,1}O_{\bullet 1}$. Thus, $ O_{\bullet 1} $ and $ M_{\ell} $ must be perfectly aligned: the valid-IV assumption ($ \tau_{\ell,0}=\infty $) pins down the first column of the orthogonal matrix $O$ up to scale, achieving point identification of the relative impulse responses to the target shock.
\end{example}

\paragraph{Relationship to existing assumptions.}
By allowing researchers to explicitly specify the proxy quality parameter $ \tau_{\ell,0} \in [0,\infty] $, our contamination assumption encompasses several popular identifying assumptions.
First, Example~\ref{eg:grr_iv} shows that the limiting case $\tau_{\ell,0} = \infty$ corresponds to the conventional valid-IV assumption \citep{mertens2013dynamic,stock2018identification}. Assuming a finite $ \tau_{\ell,0} $ then strictly weakens the exogeneity requirement.
Second, setting $\tau_{\ell,0} = 1$ recovers the ranking assumption of \cite{braun2023identification} as a special case, where the proxy must be at least as correlated with the target shock as with any other shock.
Finally, specifying $\tau_{\ell,0} = 0$ leads to the ``external variable constraints'' of \cite{ludvigson2021uncertainty}, which discard the interpretation of $m_{\ell,t}$ as a proxy for the target shock and only require $m_{\ell,t}$ to be correlated with the shock of interest.\footnote{Allowing for a positive slack term on the right-hand side of \eqref{eq:rank_O} strengthens this requirement by imposing a minimum correlation threshold without invoking any validity assumption.} These assumptions may be unnecessarily strong or weak---e.g., misspecifying $\tau_{\ell,0}=1$ for a valid IV will lead to a valid but unnecessarily wide set of responses. Our framework allows researchers to discipline the degree of contamination and conduct sensitivity analysis, as specified in Section~\ref{sec:bounds_tau}.

\subsection{Sign restrictions}
We combine the ranking restrictions with economically motivated sign restrictions. These constraints reflect the researcher's prior knowledge about either (i) the direction of specific impulse responses, or (ii) the sign of structural shocks at specific historical dates. We show that both restrictions can be cast as linear inequalities on columns of the rotation matrix $O$.

\paragraph{IRF-based sign restrictions.}
Given the definition of impulse response in Equation \eqref{eq:IRF_ijh}, restrictions of the form $\theta_{i,j,h} \ge  0$---requiring the response of variable $i$ to shock $j$ at horizon $h$ to be non-negative---translate into
\begin{equation}
    r_{i,j,h}(\phi)'\, O_{\bullet j} \ge 0,
    \qquad
    r_{i,j,h}(\phi) \coloneqq (C_h L)' e_i.
    \label{eq:sign-IRF}
\end{equation}
We collect indices of such restrictions in $\mathcal R_{\mathrm{IRF}} \subset \{1,\ldots,n\}^2 \times \{0,\ldots,H\}$.
For notational brevity, we suppress the dependence of sign restriction vectors $r_{i,j,h}$ on reduced-form parameters $\phi$.
This general formulation encompasses several standard identification schemes, as illustrated below.

\begin{example}[Self-sign normalization]\label{eg:self-sign}
    Since the rotation matrix $O$ is invariant to sign flips of the columns, the sign of a structural shock is indeterminate. Following standard practice \citep[e.g.,][]{arias2018inference}, we normalize each shock to have a positive contemporaneous effect on its own variable:
    \begin{equation}
        e_i' L\, O_{\bullet j} \ge 0, \qquad j = 1,\ldots,n.
    \end{equation}
    This corresponds to setting $r_{i,i,0} = L' e_i$ for all $i$.
\end{example}

\begin{example}[Policy-sign restriction]\label{eg:policy-sign}
    In monetary applications, it is common to require the policy rate to remain non-negative at all horizons following a contractionary monetary policy shock \citep[e.g.,][]{plagborg-moller2021local}:
    \begin{equation}
        e_1' C_h L\, O_{\bullet 1} \ge  0, \qquad h = 0,\ldots,H,
    \end{equation}
    which corresponds to $r_{1,1,h} = (C_h L)' e_1$.
\end{example}

\begin{example}[Cross-variable restrictions]\label{eg:uhlig}
    \cite{uhlig2005what} assumes that a tightening monetary shock produces (i) negative responses of prices and nonborrowed reserves (indexed by set $\mathsf{I}_P$), and (ii) positive responses of the federal funds rate (indexed by $i=1$), for horizons $h = 0,\ldots,\bar H$. This can be  recast as
    \begin{equation}
        r_{i,1,h} = -(C_h L)' e_i \quad \forall i\in \mathsf{I}_P,
        \qquad \text{and} \qquad
        r_{1,1,h} = (C_h L)' e_1,
    \end{equation}
    for all $h = 0,\ldots,\bar H$.
\end{example}

\paragraph{Narrative restrictions.}
Following \cite{antolin-diaz2018narrative}, we may constrain shock signs at specific dates. Suppose narrative evidence implies $\epsilon_{j,t_0}>0$ for some shock $j$ at date $t_0$. Using the structural mapping $\epsilon_{t} = O' L^{-1} u_{t}$, we have
\begin{equation}
    \epsilon_{j,t_0}
    = e_j' B^{-1} u_{t_0}
    = (L^{-1} u_{t_0})'\, O_{\bullet j}~.
\end{equation}
This narrative information translates into the restriction
\begin{equation}
    r_{j,t_0}(\phi)'\, O_{\bullet j} \ge  0,
    \qquad
    r_{j,t_0}(\phi) \coloneqq L^{-1} u_{t_0}.
    \label{eq:sign-event}
\end{equation}
We collect such restrictions in $\mathcal R_{\mathrm{nar}} \subset \{1,\ldots,n\} \times \mathbb T$, where $\mathbb T$ denotes the set of dates with narrative information.
Again, the dependence of $r_{j,t_0}$ on $\phi$ is suppressed when the context is clear.

\paragraph{Admissible set.}
Combining IRF and narrative restrictions, the sign-restriction-admissible set is
\begin{equation}
    \mathcal{F}_{\textup{sign}}(\phi)
    =
    \Big\{
    O\in \mathcal O(n):
    r_{i,j,h}' O_{\bullet j}\ge 0 \ \forall (i,j,h)\in \mathcal R_{\mathrm{IRF}}\, ,\;
    r_{j,t_0}' O_{\bullet j}\ge 0 \ \forall (j,t_0)\in \mathcal R_{\mathrm{nar}}
    \Big\}.
    \label{eq:O_sign}
\end{equation}
Each $O \in \mathcal{F}_{\textup{sign}}(\phi)$ generates a structural matrix $B = LO$ consistent with the imposed sign restrictions.
The associated set for the target structural vector is
\begin{equation}\label{eq:feas_set_sign_O1}
    \mathcal{G}_{\textup{sign}}(\phi)\coloneqq \{ O_{\bullet 1}: O \in \mathcal{F}_\textup{sign}(\phi) \}\subseteq \mathbb{S}^{n-1},
\end{equation}
where $\mathbb{S}^{n-1} \coloneqq \{x \in \mathbb{R}^n: \|x\| = 1\}$ denotes the unit $(n-1)$-sphere.
Throughout, we assume that the restrictions are correctly specified:
\begin{assumption}[Sign restrictions]\label{ass:sign}
    The true rotation matrix $O_0$ satisfies the imposed sign restrictions, i.e., $O_0\in\mathcal{F}_{\textup{sign}}(\phi_0)$.
\end{assumption}
Assumption~\ref{ass:sign} is standard in the sign restrictions literature \citep[e.g.,][]{uhlig2005what,baumeister2015sign,antolin-diaz2018narrative}, which ensures the feasible set $\mathcal{F}_{\textup{sign}}(\phi_0)$ is non-empty. Note that non-emptiness is necessary but not sufficient for valid identification---misspecified sign restrictions may yield non-empty sets of responses that exclude the true structural impulse response. Assumption~\ref{ass:sign} rules out such misspecification. 

\subsection{The Identified Set}
We achieve identification by combining generalized ranking restrictions with sign  restrictions.\footnote{Practitioners may not always have prior knowledge on sign restrictions. In this case, we recommend imposing self-sign restrictions only as in \autoref{eg:self-sign}, which serve as a normalization without loss of generality.} In practice, researchers rarely possess reliable prior information to calibrate proxy-specific quality parameters $\tau_{\ell,0}$.
We therefore impose a common quality parameter $\tau$ across all proxies and work with the homogeneous GRR:
\begin{equation}
    O_{\bullet 1}' M_\ell
    \;\ge\;
    \tau \,
    \big| O_{\bullet j}' M_\ell \big|\,,
    \qquad \ell = 1,\ldots,k\,,\;\; j= 2,\ldots, n \, .
    \label{eq:rank_O_homog}
\end{equation}

The feasible set depends on a collection of reduced-form parameters through \eqref{eq:O_sign} and \eqref{eq:rank_O_homog}. Since $\{ C_h \}_{h=0}^{H}$, $L$, and $ \{ M_{\ell} \}_{\ell=1}^{k} $ are continuous functions of $\phi$, the feasible set defines a correspondence in $(\tau, \phi)$.

\begin{definition}[Feasible Rotations]\label{def:feas_set}
    For a given proxy quality parameter $\tau \geq 0$ and reduced-form parameters $\phi$, the feasible set of rotation matrices is
    \begin{equation}\label{eq:feas_set}
        \mathcal{F}(\tau, \phi) \coloneqq \Big\{ O \in \mathcal{F}_{\textup{sign}}(\phi) : O_{\bullet 1}' M_\ell \geq \tau \big| O_{\bullet j}' M_\ell \big|, \quad \forall \ell = 1,\ldots,k, \; j = 2,\ldots,n \Big\}.
    \end{equation}
    The associated set for the target structural vector is
    \begin{equation}\label{eq:feas_set_O1}
        \mathcal{G}(\tau, \phi) \coloneqq \{ O_{\bullet 1}: O \in \mathcal{F}(\tau, \phi) \}\subseteq \mathbb{S}^{n-1}\,.
    \end{equation}
\end{definition}

\begin{definition}[Feasible Impulse Responses]\label{def:identified_set}
    For a given proxy quality parameter $\tau \geq 0$ and reduced-form parameters $\phi$, the identified set for the impulse response of variable $i$ to shock $\epsilon_{1,t}$ at horizon $h$ is
    \begin{equation}\label{eq:identified_set}
        \Theta_{i,h}(\tau, \phi) \coloneqq \Big\{ e_i' \, C_h(\phi)\, L(\phi)\, q : q \in \mathcal{G}(\tau, \phi) \Big\}.
    \end{equation}
    We collect these into $\Theta(\tau, \phi) \coloneqq \big\{\Theta_{i,h}(\tau, \phi) :\, i=1,\ldots,n ;\, h=0,\ldots,H\big\}$.
\end{definition}
Throughout, the dependence on $\phi$ is often suppressed and reintroduced only in the asymptotic analysis.

Under Assumptions~\ref{ass:svar}-\ref{ass:sign}, if the researcher specifies $\tau = \tau_0 \coloneqq \min_{\ell=1,\dots,k}\tau_{\ell,0}$, then the true impulse responses satisfy $\theta_{i,1,h}^0 \in \Theta_{i,h}(\tau_0)$ for all $i$ and $h$, achieving valid identification.
Moreover, the researcher's choice of $\tau$ determines the width and validity of the identified set through the following monotonicity property.

\begin{proposition}[Monotonicity]\label{prop:monotonicity}
    For any $0 \leq \tau_{\textsf{small}} < \tau_{\textsf{large}}$,
    \begin{align}
        \mathcal{F}(\tau_{\textsf{large}}) \subseteq \mathcal{F}(\tau_{\textsf{small}}) \quad \text{and} \quad \Theta(\tau_{\textsf{large}}) \subseteq \Theta(\tau_{\textsf{small}}).
    \end{align}
\end{proposition}
This proposition states that as $ \tau $ increases, the identified set shrinks. This reflects a fundamental trade-off between identifying power and robustness: Specifying $\tau > \tau_0$ yields tighter bounds but risks excluding the truth, while specifying $\tau < \tau_0$ guarantees coverage of the truth at the cost of reduced precision.

Crucially, the true proxy quality $\tau_0$ is not identified without further assumptions---analogous to the impossibility of testing instrument validity using outcome data alone. We address this through two complementary approaches. Section~\ref{sec:bounds_tau} exploits complementarities between the proxy zoo and sign restrictions to partially identify an outer set $[0, \overline{\tau}]$ containing $\tau_0$. Section~\ref{sec:sensitivity} conducts sensitivity analysis by reporting $\Theta_\tau$ as $\tau$ varies over this plausible range, following established practice in partial identification \citep{andrews2017measuring,vanderweele2017sensitivity,masten2020inference}.

\subsection{Implementation}\label{sec:implementation}

We now describe how to compute the identified set in practice.
Let $\hat{\phi}$ denote a consistent estimator of the reduced-form parameters $\phi$ defined in \eqref{eq:def_phi}, obtained via OLS for the VAR coefficients, $(\hat A_1,...,\hat A_p)$, and sample moments for $\Sigma$ and $\{M_\ell\}_{\ell=1}^k$, i.e.,
\begin{equation*}
    \hat{\Sigma}=\frac{1}{T-p}\sum_{t=p+1}^{T} \hat{u}_t \hat{u}_t', \quad \hat{M}_\ell = \frac{1}{T-p}\sum_{t=p+1}^{T} \hat{L}^{-1} \hat{u}_t m_{\ell,t}\,
\end{equation*}
where $\hat L=\mathrm{Chol}(\hat\Sigma)$ the Cholesky factor of the residual covariance matrix.
Further denote $\hat C_h$ the reduced-form impulse responses in \eqref{eq:def_reduced_irf} and $\hat r_{i,j,h}:= r_{i,j,h}(\hat{\phi})$ the sample analogues of the sign-restriction vectors $r_{i,j,h}$.

Given a proxy quality parameter $\tau$, the identified set $\Theta_{i,h}(\tau) = \Big[\underline{\theta}_{i,1,h}(\tau),\, \overline{\theta}_{i,1,h}(\tau)\Big]$ is estimated by solving constrained optimization problems over the orthogonal group. The lower bound is
\begin{align}
    \hat{\underline{\theta}}_{i,1,h}(\tau) = \inf_{O \in \mathcal{O}(n)} \; e_i' \hat{C}_h \hat{L} O_{\bullet 1} \label{eq:optim_obj}
\end{align}
subject to
\begin{align}\label{eq:constraints}
    \begin{array}{lll}
        \hat{r}_{i,j,h}' O_{\bullet j} \geq  0,                                        & \forall (i,j,h) \in \mathcal{R}_{\mathrm{IRF}}     & \text{(IRF restrictions)}       \\[0.5ex]
        \hat{r}_{j,t_0}' O_{\bullet j} \geq  0,                                        & \forall (j,t_0) \in \mathcal{R}_{\mathrm{nar}}     & \text{(narrative restrictions)} \\[0.5ex]
        O_{\bullet 1}' \hat{M}_\ell \geq \tau \big| O_{\bullet j}' \hat{M}_\ell \big|, & \forall \ell = 1,\ldots,k, \; j = 2,\ldots,n \quad & \text{(ranking restrictions)}
    \end{array}
\end{align}
The upper bound $\hat{\overline{\theta}}_{i,1,h}(\tau)$ is obtained by replacing is obtained by replacing $\inf$ with $\sup$ in \eqref{eq:optim_obj}, subject to the same constraints \eqref{eq:constraints}.
The optimization can be efficiently solved using off-the-shelf nonlinear programming solvers such as \textsf{fmincon} in MATLAB; see Appendix~\ref{app:computation} for computational details.

\section{Bounds on Proxy Quality}\label{sec:bounds_tau}

The generalized ranking restrictions \eqref{eq:rank_O_homog} depend on a proxy quality parameter $\tau$ that governs a sharp trade-off: larger values tighten identified sets but impose stronger assumptions. Valid identification requires the researcher to specify $\tau \leq \tau_0$, where $\tau_0$ is the true (unknown) proxy quality. 

Rather than calibrating $\tau$ ex ante, we exploit two sources of falsifiable information to derive an empirically grounded upper bound $\bar{\tau}$ such that $\tau_0 \in [0,\bar{\tau}]$.
First, when proxies contradict IRF-based sign or narrative restrictions, this conflict reveals that proxies cannot be arbitrarily strong instruments relative to their contamination. Second, when multiple proxies provide conflicting directional information, their joint feasibility restricts admissible values of $\tau_0$. Together, these sources of disagreement discipline the quality parameter without requiring parametric assumptions about the endogeneity mechanism.

The next result formalizes this partial identification approach. Let $\alpha_{\ell,v}$ denote the angle between the proxy moment vector $M_\ell$ and a generic conformable vector $v$.

\begin{theorem}[Partial Identification of Proxy Quality]\label{thm:tau_identification}
    Under Assumptions~\ref{ass:svar}--\ref{ass:sign}, the true proxy quality parameter satisfies $\tau_0 \le \bar{\tau}$, where
    \begin{equation}\label{eq:tau_bound_tight}
        \bar{\tau}
        \coloneqq
        \sqrt{n-1}\,
        \cot\!\left(
        \min_{q \in \mathcal{G}_{\textup{sign}}}
        \max_{\ell=1,\ldots,k}
        \alpha_{\ell,q}
        \right).
    \end{equation}
\end{theorem}

Theorem~\ref{thm:tau_identification} quantifies the degree of conflicting information within the complete set of identification assumptions: the generalized ranking restrictions encoded in the proxy moment vectors $\{ M_{\ell} \}_{\ell=1}^{k}$ and the sign 
restrictions represented by $ \mathcal{G}_{\textup{sign}} $.
The bound $\bar{\tau}$ tightens when this conflict intensifies, revealing that high proxy quality is incompatible with the data and maintained restrictions.

To build intuition, recall from Example~\ref{eg:grr_iv} that perfect exogeneity assumption ($\tau_0 = \infty$) requires the first column of the rotation matrix to align exactly with the proxy moment: there must exist $ q \in \mathcal{G}_{\textup{sign}} $ such that $ q \propto M_\ell$. This alignment condition clarifies why disagreement---either between proxies and sign restrictions, or among multiple proxies---forces finite bounds on proxy quality, in a manner analogous to overidentification in classical IV settings \citep[e.g.,][]{sargan1958estimation,masten2021salvaging}.

\paragraph{Single proxy, consistent with sign restrictions.} Consider first a single proxy ($k=1$) whose moment vector $M_1$ lies within the sign-feasible set, so that $\min_{q \in \mathcal{G}_{\textup{sign}}} \alpha_{1,q} = 0$.\footnote{Without loss of generality, we consider proxy covariance vectors normalized to unit length.} Since some $q \in \mathcal{G}_{\textup{sign}}$ aligns with $M_1$ (e.g., $ q=M_1 $), the alignment poses no contradiction with the sign restrictions, and thus the data permit $ \tau_0 $ to be arbitrarily large. This is reflected by $ \bar{\tau}=\sqrt{n-1}\cot(0)=\infty $.

\paragraph{Single proxy conflicting with sign restrictions.} When the proxy moment vector lies outside the sign-feasible cone---so that $\min_{q \in \mathcal{G}{\textup{sign}}} \alpha_{1,q} > 0$---perfect alignment must violate the sign restrictions. This follows from the fact that every $ q \propto M_1 $ (under the assumption of $\tau_0 = \infty$) does not lie within the sign-restricted set $\mathcal{G}_{\textup{sign}}$. This inconsistency forces $\tau_0 < \infty$: the proxy must contain some contamination for the restrictions to remain jointly feasible. The larger the angular distance, the tighter the bound.

\paragraph{Multiple conflicting proxies.} Consider a proxy zoo with $k\geq 2$. Even when all proxies are individually consistent with sign restrictions ($M_{\ell}\in \mathcal{G}_{\textup{sign}}$ for all $ \ell=1,\ldots,k $), disagreement among proxies disciplines $\tau_0$. If $ M_{\ell}\neq M_{s} $ for some pair of proxies $\ell$ and $s$, then any $ q \propto M_{\ell} $ must satisfy $ q \not\propto M_{s} $, implying contamination in at least the $s$-th proxy. Therefore, when proxies point in different directions, their joint feasibility imposes a finite upper bound on proxy quality.

A visual illustration of these cases is given by \autoref{fig:tau_bar} in the appendix.

\subsection{Point Identification via Proxy Complementarity.} \label{sec:point_id}
The complementarity mechanism described above has a striking theoretical implication: with sufficiently complementary proxies, the collective identifying power can shrink the feasible set to a single point. In this limiting case, the zoo achieves point identification without requiring any individual proxy to be a valid instrument.

\begin{theorem}[Point Identification]\label{thm:point_identification}
    For any $\tau \geq 1$ and any $k \geq 2$, there exist proxy-covariance vectors $M_1, \ldots, M_k$ such that $\mathcal{G}(\tau)$ and thus $\Theta(\tau)$ are singletons.
\end{theorem}

Theorem~\ref{thm:point_identification} illustrates that point identification is possible purely through an informational conflict between contaminated sources: in the ideal case, two complementary proxies suffice for point identification. While perfect complementarity is unlikely in practice, this result highlights complementarity as an alternative channel for sharp identification to exogeneity assumptions.

\begin{remark}
    Throughout the section, the identifying power of proxies depends on their covariance vectors $\{M_\ell\}_{\ell=1}^k$ rather than the raw variables $\{m_{\ell,t}\}$.
    To see this, recall that $ M_{\ell} = \mathbb{E}[L^{-1}u_{t} m_{\ell,t}] $ measures how proxy $m_{\ell,t}$ responds to orthogonalized reduced-form errors $\tilde{u}_t \coloneqq L^{-1}u_t$.\footnote{This is because $ M_{\ell} = \mathbb{E}[\tilde{u}_{t} m_{\ell,t}] = \left( \mathbb{E}[\tilde{u}_{t}\tilde{u}_{t}'] \right)^{-1} \mathbb{E}[\tilde{u}_{t} m_{\ell,t}] $.}
    When $M_\ell$ closely aligns with the reduced-form IRFs $C_h L$, the proxy and policy variables respond similarly to reduced-form shocks, reflecting a lack of \textit{external} information about the target shock.
    As a result, even if two proxies are uncorrelated with each other---a tempting indicator of complementarity---they provide redundant identifying information if their covariance vectors $M_\ell$ point in similar directions, meaning both respond similarly to the orthogonalized errors $ L^{-1}u_{t}$.
\end{remark}

\section{Sensitivity analysis: a diagnostic framework}\label{sec:sensitivity}
Having established the upper bound for the proxy quality $ \tau_0 \in [0,\overline{\tau}] $, this section develops formal tools for sensitivity analysis and diagnostic assessment of the proxy zoo.
We first characterize breakdown values that demarcate when specific substantive conclusions become supported by the data \citep[e.g.,][]{vanderweele2017sensitivity,masten2020inference}. We then introduce proxy diagnostics to evaluate the information content of individual proxies relative to the zoo. Finally, we provide practical guidelines for implementing sensitivity analysis in empirical work.

\subsection{Breakdown values}
Researchers often seek to establish specific substantive conclusions---such as whether a monetary policy shock raises output, or the magnitude of the effect exceeds a certain threshold. In the current framework, such conclusions may or may not hold as the proxy quality assumption $ \tau_0 $ varies. We formalize this sensitivity through the concept of breakdown value.

\begin{definition}[Breakdown value]\label{def:breakdown_val}
    For a claim of interest $c: \mathbb{R}^{n\times (H+1)} \to \{0,1\}$ about the impulse response $\theta \in \Theta(\tau,\phi)$, the breakdown value is
    \begin{equation}\label{eq:breakdown_val}
        \tau^{\ast}(c,\phi) \coloneqq \inf\Big\{\tau \geq 0: c(\theta) = 1 \text{ for all } \theta \in \Theta(\tau,\phi)\Big\}.
    \end{equation}
    When clear from context, we suppress the dependence on $\phi$.
\end{definition}

The breakdown value $\tau^{\ast}(c)$ represents the \textit{weakest} possible proxy quality assumption required to support the claim $c$. If the researcher believes $\tau_0 \geq \tau^{\ast}$, then the claim $c$ is consistent with the identified set; otherwise, the claim lacks support.

We illustrate popular claims through the following examples. Throughout, we denote by $\mathcal{H}$ some horizon set and without loss of generality consider generic variable indices $ i $ and $ j $. Moreover, we suppress the dependence of the breakdown value $ \tau^{\ast} $ on claim-specific parameters such as the horizon set $ \mathcal{H} $.

\begin{example}[Sign claims]\label{ex:breakdown_sign}
    For the claim $c(\theta) = \mathbf{1}\{\theta_{i,h} \ge   0\text{ for } h\in \mathcal{H}\}$ (positive response), the breakdown value is
    \begin{equation}
        \tau_{\textup{pos}}^{\ast}(c) = \inf\{\tau \geq 0: \underline{\theta}_{i,h}(\tau) \ge  0\text{ for } h\in \mathcal{H}\}.
    \end{equation}
    The claim may also involve responses across variables. For example, the breakdown value for the claim $c(\theta) = \mathbf{1}\{\theta_{i,h} \ge   0,\, \theta_{j,h}\le 0\text{ for } h\in \mathcal{H}\}$ is
    \begin{equation}
        \tau_{\textup{sign}}^{\ast}(c) = \inf\{\tau \geq 0: \underline{\theta}_{i,h}(\tau) \ge  0,\, \overline{\theta}_{j,h}\le 0\text{ for } h\in \mathcal{H}\}.
    \end{equation}
\end{example}

\begin{example}[Magnitude claims]\label{ex:breakdown_magnitude}
    Consider the claim $c(\theta) = \mathbf{1}\{ \theta_{i,h} \geq \delta\}$ with some threshold $\delta > 0$,
    \begin{equation}
        \tau_{\textup{mag}}^{\ast}(c) = \inf\{\tau \geq 0: \underline{\theta}_{i,h}(\tau) \geq \delta\}~.
    \end{equation}
\end{example}

\paragraph{Breakdown frontiers.} For multiple claims or claims involving additional parameters---such as the horizon set $ \mathcal{H} $---it is useful to visualize the \textit{breakdown frontier} as the locus of parameter pairs where conclusions change. The frontier partitions the assumption/conclusion space into regions where different conclusions are supported, providing a comprehensive sensitivity map.

\subsection{Proxy diagnostics}\label{sec:proxy_diag}
When multiple proxies are available, researchers naturally ask: Which proxies are most informative? Is the proxy zoo delivering genuine complementarity, or are some proxies redundant? How sensitive are the conclusions to specific proxies? We develop a set of diagnostic tools to answer these questions.

We start by defining the informativeness of a proxy zoo. Formally, let $\mathcal{M}$ be a set of proxies and denote the identified set obtained using this proxy set as $\Theta_{i,h}(\mathcal{M},\tau)$. We further denote the identified set using sign and narrative restrictions only by
\begin{equation}\label{eq:sign_only_set}
    \Theta_{i,h}^{\textup{sign}}(\phi) \coloneqq \Big\{ e_i' , C_h(\phi), L(\phi), q : q \in \mathcal{G}_\textup{sign}(\phi) \Big\}~.
\end{equation}
We then measure the information content of the proxy zoo $\mathcal{M}$ as follows.

\begin{definition}[Proxy Zoo Information]\label{def:zoo_information}
    At quality level $\tau$, the information of a proxy zoo relative to sign and narrative restrictions is measured by
    \begin{equation}\label{eq:zoo_information}
        \kappa(\mathcal{M},\tau) = \frac{1}{n(H+1)}\sum_{i=1}^n \sum_{h=0}^H \kappa_{i,h}(\mathcal{M},\tau),\quad \kappa_{i,h}(\mathcal{M},\tau) \coloneqq 1-\frac{\text{width}(\Theta_{i,h}(\mathcal{M},\tau))}{\text{width}(\Theta_{i,h}^{\textup{sign}})}.
    \end{equation}
\end{definition}
Intuitively, $\kappa_{i,h}(\mathcal{M},\tau)$ captures the reduction in the width of the identified set achieved by using the proxy set $\mathcal{M}$, expressed as a fraction of the baseline width implied by sign restrictions alone.
At the population level, the width of the identified set is non-negative and weakly decreases when a proxy zoo is introduced; hence, $\kappa(\mathcal{M},\tau) \in [0,1] $ by construction. A value of $\kappa(\mathcal{M},\tau)$ close to 1 indicates that the proxy zoo $\mathcal{M}$ substantially tightens the identified set, while $\kappa(\mathcal{M},\tau) \approx 0$ suggests that the zoo provides little information beyond the sign restrictions.
In practice, the measure may slightly exceed 1 because standard numerical solvers (e.g., \textsf{fmincon} in MATLAB) are not guaranteed to attain global optima when computing the bounds of $\Theta_{i,h}(\mathcal{M},\tau)$.

\paragraph{Information Analysis.}
Given Definition~\ref{def:zoo_information}, we can directly compare the information content across different proxy zoos. For two generic proxy zoos $\mathcal{M}$ and $\widetilde{\mathcal{M}}$, define the relative informativeness of zoo $\mathcal{M}$ compared with $\widetilde{\mathcal{M}}$ as
\begin{equation}\label{eq:relative_informativeness}
    \Delta(\mathcal{M},\widetilde{\mathcal{M}},\tau) \coloneqq \kappa(\mathcal{M},\tau) - \kappa(\widetilde{\mathcal{M}},\tau).
\end{equation}
For any nested proxy zoos, i.e., $\widetilde{\mathcal{M}}\subset \mathcal{M}$, theoretical monotonicity implies $\kappa(\mathcal{M},\tau) \ge \kappa(\widetilde{\mathcal{M}},\tau)$.\footnote{In finite-sample implementations, numerical solvers used to compute the bounds of
    $\Theta_{i,h}(\mathcal{M},\tau)$ may converge to local optima, leading to mild violations of monotonicity. We interpret such cases as indicating a negligible marginal contribution from the additional proxies in
    $\mathcal{M}\setminus\widetilde{\mathcal{M}}$.}
For non-nested proxy zoos, $\Delta(\mathcal{M},\widetilde{\mathcal{M}},\tau)$ provides a purely comparative measure without a monotonic interpretation.

\paragraph{Leave-one-proxy-out (LOPO) Information Analysis.} A particularly useful special case is the leave-one-proxy-out (LOPO) analysis. Let $\mathcal{M}^{(-\ell)}$ denote the set of all proxies in the zoo $\mathcal{M}$ excluding proxy $\ell$. The informativeness of the $\ell$-th proxy is quantified by
\begin{equation}\label{eq:lopo_informativeness}
    \Delta_{\ell}(\tau) \coloneqq \kappa(\mathcal{M},\tau) - \kappa(\mathcal{M}^{(-\ell)},\tau).
\end{equation}
This measure captures the marginal contribution of proxy $\ell$ to tightening the identified set. A large value of $\Delta_\ell(\tau)$ indicates that proxy $\ell$ is highly informative conditional on the remaining proxies, while values close to zero indicate redundancy.

\subsection{Empirical guidelines}
We conclude with a practical workflow for implementing sensitivity analysis in applied work.
\paragraph{Step 1: Establish the feasible range.}
Compute the upper bound $\bar{\tau}$ from Theorem~\ref{thm:tau_identification}. This defines the plausible range $\tau \in [0, \bar{\tau}]$ over which to conduct sensitivity analysis.\footnote{If $\bar{\tau} = \infty$, use a large $ \overline{\tau} $ for the following sensitivity analysis.}

\paragraph{Step 2: Compute identified sets on a grid.}
For a fine grid $\{\tau_1, \ldots, \tau_G\} \subset [0, \bar{\tau}]$, solve the optimization problems in \eqref{eq:optim_obj}--\eqref{eq:constraints} to obtain $\{\Theta_{i,h}(\tau_g,\hat{\phi})\}_{g=1}^G$. Plot the bounds $[\underline{\theta}_{i,h}(\tau,\hat{\phi}), \overline{\theta}_{i,h}(\tau,\hat{\phi})]$ as functions of $\tau$ for target variables $i$ and horizons $h$.

\paragraph{Step 3: Report breakdown frontiers.}
For substantive conclusions of interest (e.g., ``output rises on impact'', ``price response is negative''), compute and report $\tau^{\ast}(c)$ using Definition~\ref{def:breakdown_val}. Interpret these as ``minimal contamination tolerance'' required to support each conclusion.

\paragraph{Step 4: Conduct proxy diagnostics.}
Compute proxy zoo informativeness $\kappa$ as in \eqref{eq:zoo_information} and LOPO information measures $\{\Delta_\ell(\tau)\}_{\ell=1}^k$ as in \eqref{eq:lopo_informativeness}. Report the sensitivity of conclusions to proxies in the zoo.

\paragraph{Step 5: Compare to point-identified benchmarks.}
When a particular proxy is of interest, e.g., treated as a valid IV in the benchmark model, compute the point-identified impulse responses under $\tau = \infty$. Compare it to the robust bounds under finite $\tau \in [\tau^{\ast}(c),\overline{\tau}]$. This quantifies the cost of relaxing exogeneity and highlights where conclusions are robust versus fragile.

\section{Simulation Study}\label{sec:simulation}
In this section, we illustrate the proposed framework using a simulation study based on a medium-scale dynamic stochastic general equilibrium (DSGE) model. We generate data from the model of \citet{smets2007shocks} for $T=5000$ periods and estimate a VAR$(12)$ on seven observable variables: output growth, consumption growth, investment growth, wage growth, inflation, the nominal interest rate, and employment.\footnote{We consider a large sample size to isolate identification challenges from finite-sample uncertainty.}

A common practice in the empirical literature is to recover structural shock series using set-identified methods---most notably sign restrictions---and subsequently use the recovered shock series as external proxies in further analyses \citep[e.g.,][]{baumeister2019structural,jarocinski2020deconstructing}. To mirror this practice, we first identify the monetary policy shock using standard sign restrictions in the spirit of \citet{uhlig2005what}. Specifically, we impose correct sign restrictions on the responses to a contractionary monetary policy shock for up to two horizons, leaving the response of output growth unrestricted.

\begin{figure}[ht]
    \centering
    \includegraphics[width=\textwidth]{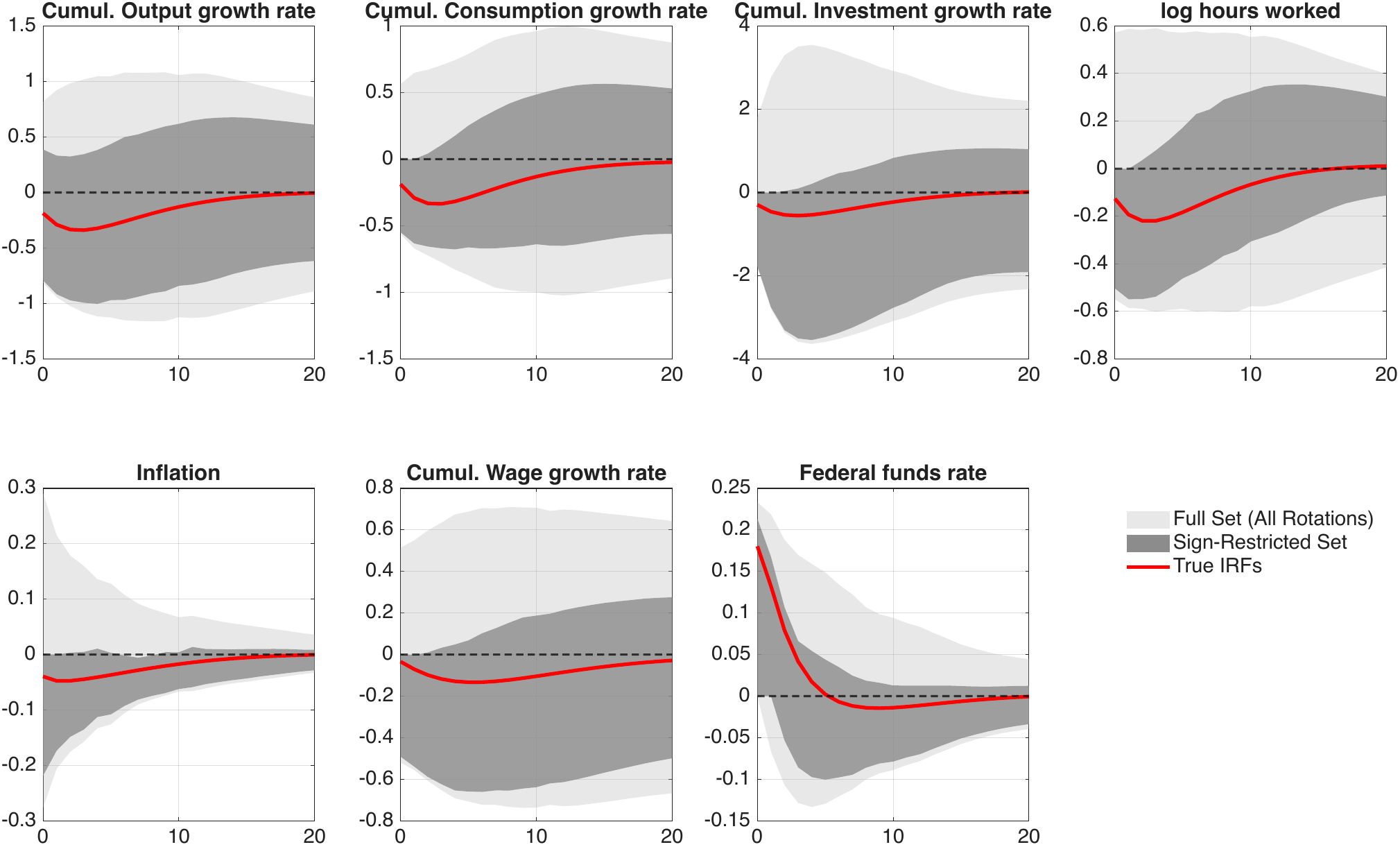}
    \caption{Identified sets of impulse responses under sign restrictions.}
    \label{fig:SIM-SW-UhligSet}
    \begin{fignote}[\linewidth]
        \emph{Notes:} The red line reports the true impulse response to a monetary policy shock from the \citet{smets2007shocks} model. Shaded areas denote identified sets under the sign restrictions.
    \end{fignote}
\end{figure}

\autoref{fig:SIM-SW-UhligSet} reports the identified set under the above sign restrictions. Perhaps surprisingly, the identified sets of output responses remain large and relatively uninformative, although the sign restrictions do reduce the set of admissible impulse responses. Moreover, the identified sets contain the true impulse responses, which is consistent with the implication of Proposition~\ref{prop:monotonicity} when GRR are absent.

Given the set of admissible structural representations, we construct proxy series for the monetary policy shock following two standard approaches. First, we define a \emph{Median-$B$} proxy, obtained from the structural shock series associated with the element-wise median of admissible contemporaneous impact matrices. Second, we define a \emph{Closest-$B$} proxy, corresponding to the admissible impact matrix that is closest (in Euclidean distance) to the element-wise median.

\begin{figure}[ht]
    \centering
    \includegraphics[width=\linewidth]{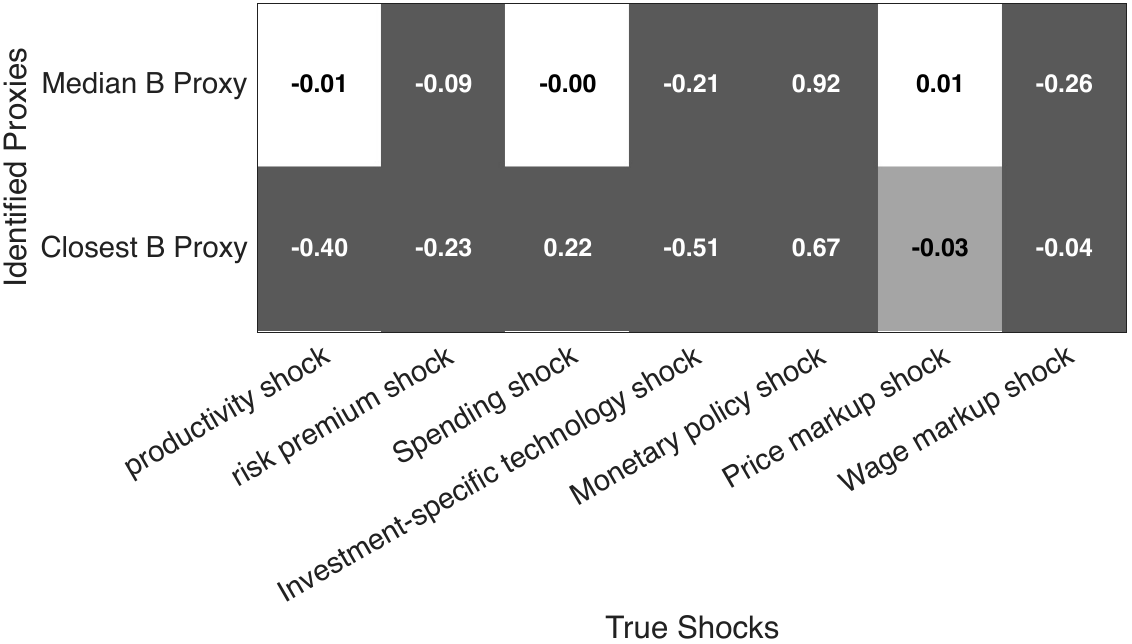}
    \caption{Correlation significance map of sign-restricted shock}
    \begin{fignote}[\linewidth]
        \emph{Notes:} Monetary policy proxies constructed using \cite{uhlig2005what}'s sign restrictions vs.\ true structural shocks in \cite{smets2007shocks}. Light gray: $p\text{\normalfont-val}\leq 0.10$; gray: $p\text{\normalfont-val}\leq 0.05$; dark gray: $p\text{\normalfont-val}\leq 0.01$.
    \end{fignote}
    \label{fig:SIM-SW-Uhlig-Heatmap}
\end{figure}

While both proxies are consistent with the imposed sign restrictions, they are not guaranteed to be orthogonal to other structural shocks in the data generating process. \autoref{fig:SIM-SW-Uhlig-Heatmap} illustrates this point by reporting correlations between the constructed proxies and the \emph{true} structural shocks of the DSGE model. Even with the very large sample size, both proxies exhibit non-negligible correlations with non-monetary shocks, including technology and wage markup shocks.\footnote{With realistic sample sizes, such contamination becomes even more prominent.} Moreover, there is no systematic pattern indicating which non-target shocks are more likely to load on the proxy, nor which construction method yields a uniformly cleaner proxy.

\begin{figure}[ht]
    \centering
    \includegraphics[width=\textwidth]{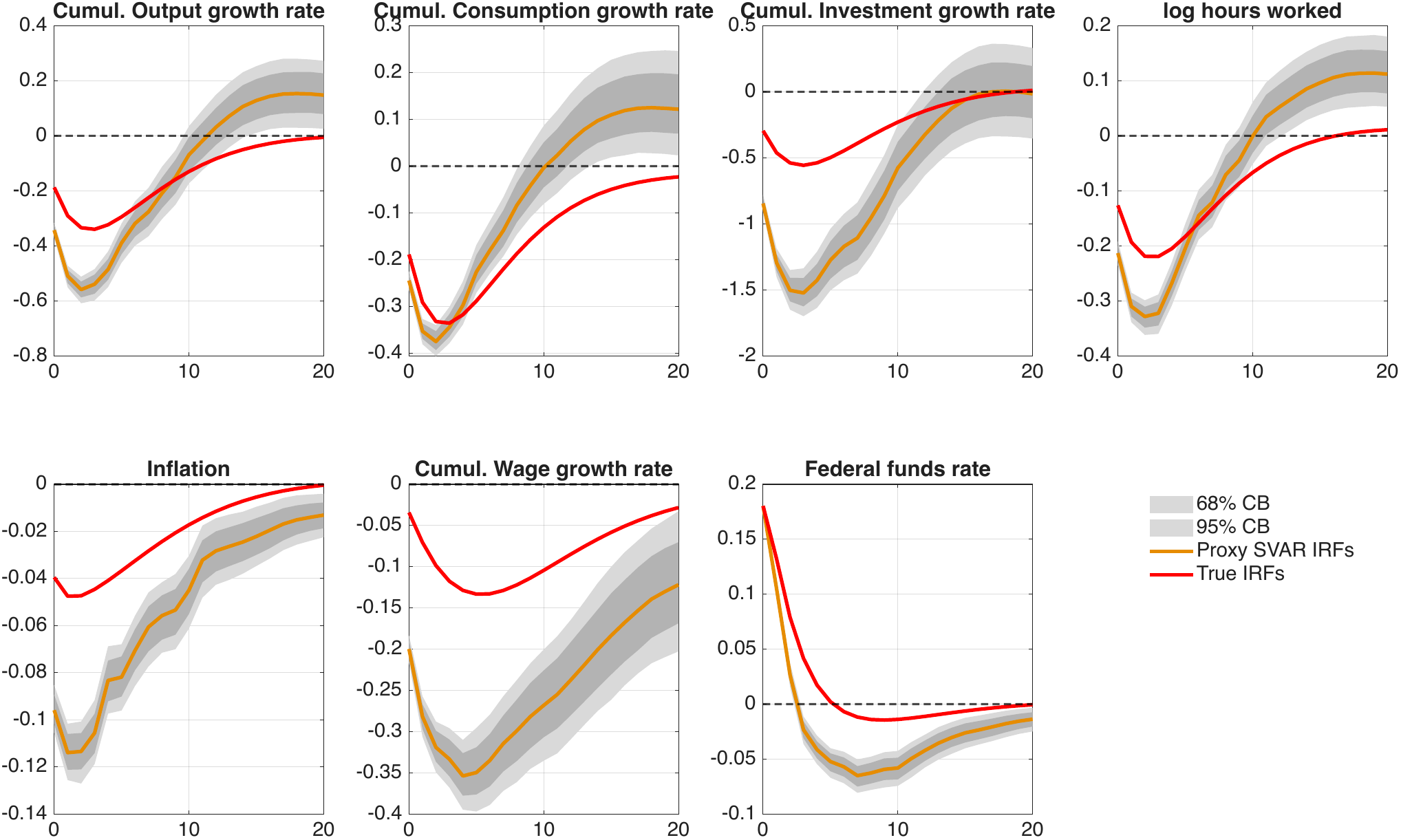}
    \caption{Bias in Median-$B$ Proxy-SVAR Estimates}
    \begin{fignote}[\linewidth]
        \emph{Notes:} The red line reports the true impulse response to a monetary policy shock from the \citet{smets2007shocks} model; the orange line reports the Median-$B$ proxy-SVAR estimate. Gray shaded areas denote pointwise moving block bootstrap (MBB) confidence intervals \citep{jentsch2022asymptotically}.
    \end{fignote}
    \label{fig:SIM-SW-ProxySVAR}
\end{figure}

To illustrate the failure of point identification given contaminated proxies, we follow \cite{mertens2013dynamic} and estimate the structural impulse responses using the Median-$ B $ proxy. As \autoref{fig:SIM-SW-ProxySVAR} shows, the resulting impulse response estimates are severely biased. Moreover, the signs happen to be consistent with the truth, making it difficult to detect identification failure from conventional ``puzzling'' responses alone.

We next apply the GRR, following Section~\ref{sec:implementation}, to characterize identification in the presence of such contamination. Rather than imposing exact orthogonality, we consider relaxation sets indexed by the proxy quality parameter $\tau$, which bounds the proxy's covariance with non-target shocks relative to the target shock. Identified sets are computed over a grid $\tau \in [0,20]$. The true value $\tau_0=3.5438$, computed from the simulation metadata, is used only for validation and is not observable in applications.

\begin{figure}[ht]
    \centering
    \includegraphics[width=\textwidth]{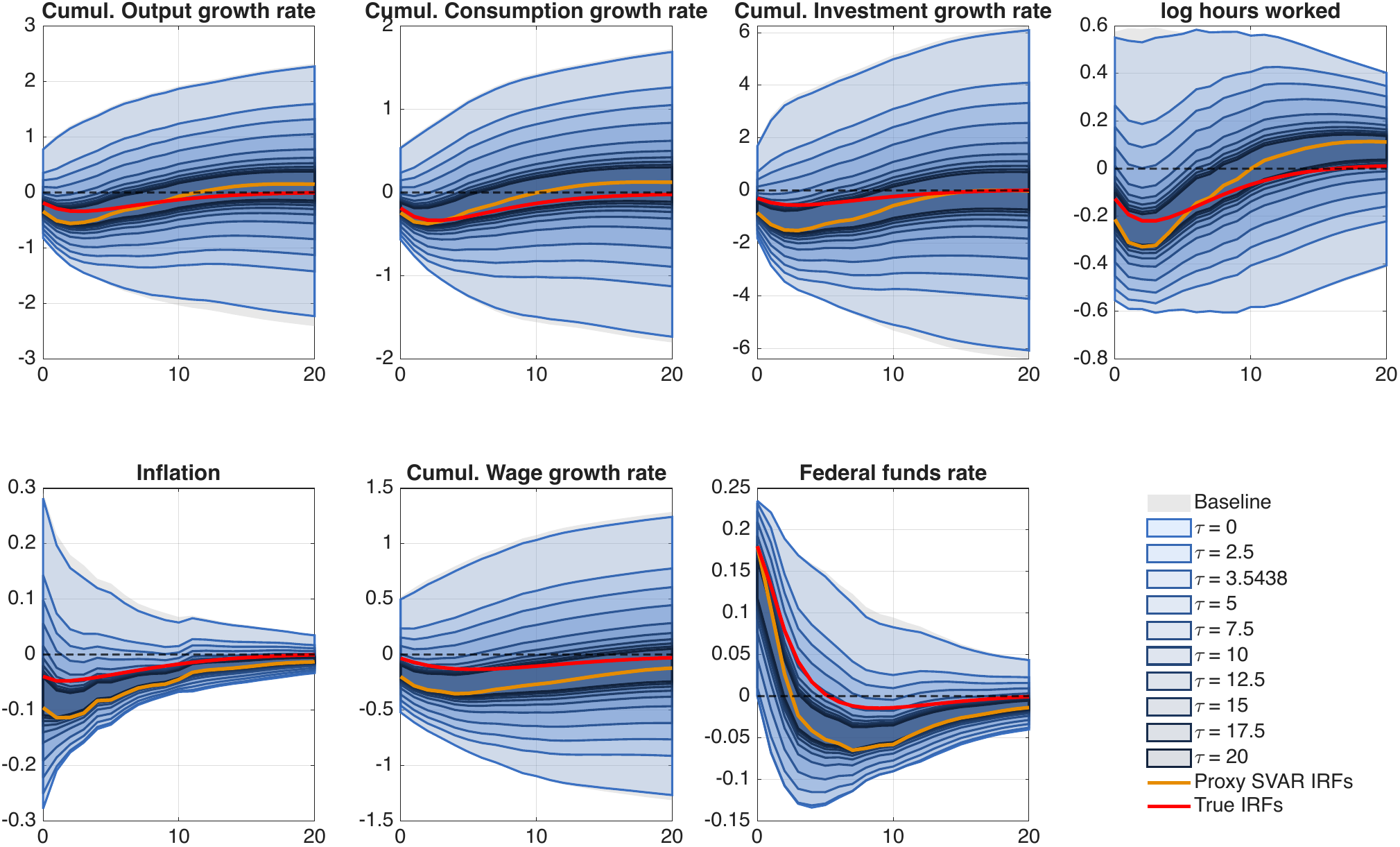}
    \caption{Set-identified impulse responses under GRR.}
    \begin{fignote}[\linewidth]
        \emph{Notes:} The red line reports the true impulse response to a monetary policy shock from the \citet{smets2007shocks} model; the orange line reports the Median-$B$ proxy-SVAR estimate. Each shaded area corresponds to the identified set for a specific $ \tau $.
    \end{fignote}
    \label{fig:SIM-SW-Proxy1-Sensitivity}
\end{figure}

\autoref{fig:SIM-SW-Proxy1-Sensitivity} reports the identified sets of impulse responses obtained using the Median-$B$ proxy. The red line shows the true impulse response, while the orange line corresponds to the point-identified proxy-SVAR that treats the proxy as exactly exogenous.\footnote{In population, the point-identified IRF that treats the contaminated proxy as exactly exogenous is given by Equation \eqref{eq:cov_contam}}

Two patterns emerge. First, for a wide range of values of $\tau$, the identified sets contain the true impulse response, which is consistent with Proposition~\ref{prop:monotonicity}. As $\tau$ increases and stronger assumptions are imposed, the identified sets shrink toward the point-identified proxy-SVAR estimate. Second, the sensitivity analysis shows that informative and robust inference is possible without imposing exact exogeneity of the proxy.

This experiment highlights that proxies constructed from set-identified procedures need not satisfy exact orthogonality with non-target shocks, even in large samples. The proposed framework provides a formal and transparent way to conduct identification in such environments.

Finally, we note a sharp contrast in computational burden between the simulation and empirical settings. The simulation relies on repeated sampling over admissible rotations, whereas in the empirical application with eight proxies the identified sets are computed via deterministic constrained optimization, as in~\eqref{eq:optim_obj}. For a fixed value of $\tau$ (for example, at a breakdown value), computing impulse-response bounds takes approximately seven seconds on a standard machine, indicating that the proposed approach is computationally tractable even with a rich proxy zoo.

\section{Applications}\label{sec:application}
In this section, we apply the proposed framework to revisit the effects of monetary policy shocks. We identify the set of impulse responses using the eight proxies described in Section~\ref{sec:motivation}, imposing self-sign restriction in \autoref{eg:self-sign} and the generalized ranking restrictions in Equation~\eqref{eq:rank_O}. Since the self-sign restriction serves only to resolve sign indeterminacy, identification relies entirely on the information contained within the set of proxies (the ``proxy zoo'').

\begin{figure}[ht]
    \centering
    \includegraphics[width=1\linewidth]{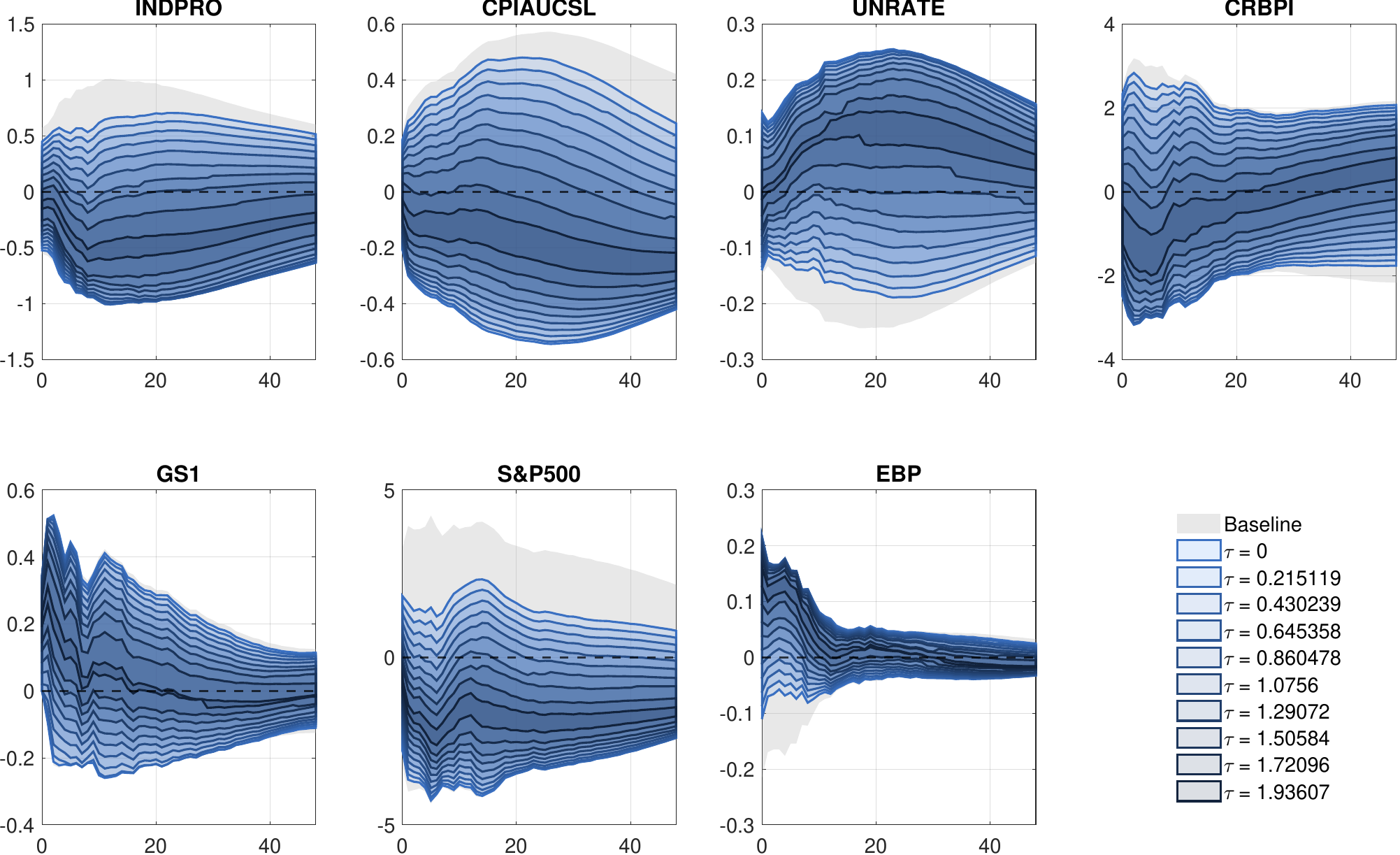}
    \caption{Proxy Zoo Identification: Sensitivity Analysis}
    \begin{fignote}[\linewidth]
        \emph{Notes:} Sensitivity of impulse responses to a contractionary monetary policy shock. The IRFs are identified using the self-sign normalization in \autoref{eg:self-sign} and the generalized ranking restrictions in Equation~\eqref{eq:rank_O}, utilizing all eight proxies listed in Section~\ref{sec:motivation}. Each shaded area corresponds to the identified set for a specific $ \tau $. The gray area depicts the identified set under self-sign normalization only.
    \end{fignote}
    \label{fig:Sensitivity}
\end{figure}

Following our empirical guidelines, we first report the identified impulse responses over a grid of proxy quality parameters $ \tau \in [0,\overline{\tau}) $ in \autoref{fig:Sensitivity}.
Three key insights emerge from this analysis.
First, leveraging the proxy zoo allows us to partially identify the upper bound of the proxy quality parameter, $ \overline{\tau}=2.15 $. This result implies that at least some proxies in the set violate the strict exogeneity assumption, corroborating recent skepticism regarding proxy validity \citep[e.g.,][]{bruns2024testing}.
Second, under relatively conservative assumptions regarding proxy quality---for example, $ \tau=0 $ (corresponding to external variable restrictions, as in \citet{ludvigson2021uncertainty}) or $ \tau=1 $ (corresponding to ranking restrictions, as in \citet{braun2023identification})---the identified sets remain wide and uninformative, despite the inclusion of multiple proxies.
Third, as we strengthen the identification assumption by increasing $ \tau $, the identified sets shrink as expected. Notably, the identified sets become informative at finite values of $ \tau $.

We formalize this intuition through a breakdown value analysis. Suppose we postulate that industrial production and inflation must fall contemporaneously following a monetary tightening. Our analysis reveals that this claim holds only if we impose $ \tau\ge  \tau ^{\ast}= 1.88 $. Intuitively, this requires the proxies to be, on average, approximately $ 1.9 $ times more correlated with the monetary policy shock than with any other structural shock.
Collectively, these results suggest that for applied researchers, the standard assumption of valid IVs (infinite $ \tau $) may be excessively strong and is unnecessary for establishing certain qualitative conclusions.

Next, we examine the sensitivity of these empirical claims to the composition of the proxy set. To this end, we conduct a leave-one-proxy-out (LOPO) analysis (see Section~\ref{sec:proxy_diag}), fixing the proxy quality parameter at the breakdown value $ \tau^{\ast} $. Recall that any assumption weaker than $ \tau^{\ast} $ is insufficient to support the claim that both industrial production and inflation contract following the shock.\footnote{Importantly, a wide identified set does not necessarily imply that the claims of interest are \emph{rejected}; rather, it indicates that there is insufficient information to draw a definitive conclusion.} By excluding one proxy at a time, we assess the stability of this claim and the informativeness of individual proxies.

\begin{figure}[h!]
    \centering
    \includegraphics[width=\linewidth]{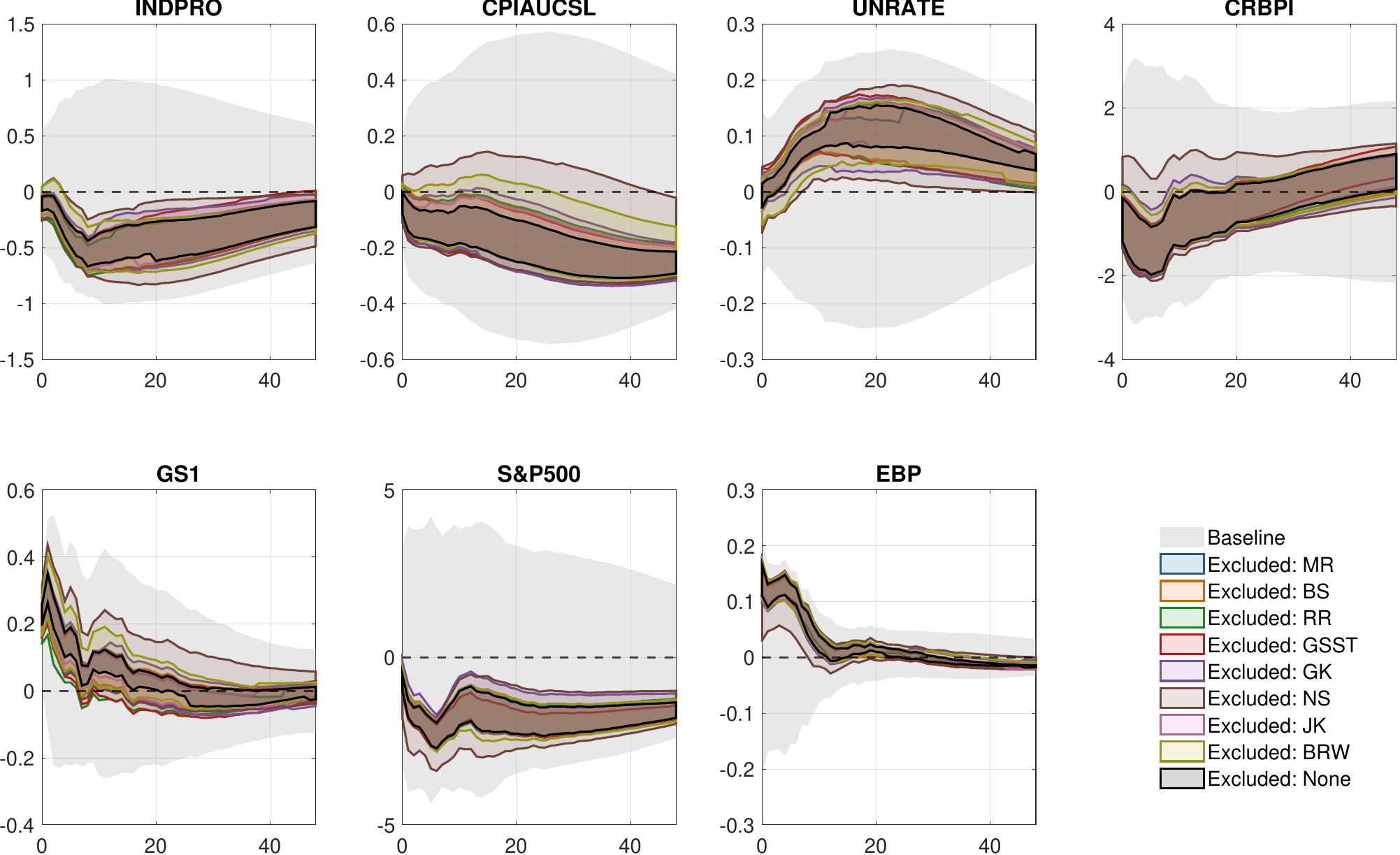}
    \caption{Proxy Zoo Identification: Leave-One-Proxy-Out Analysis}
    \begin{fignote}[\linewidth]
        \emph{Notes:} Sensitivity of impulse responses to a contractionary monetary policy shock. The identification scheme follows \autoref{fig:Sensitivity}. Each shaded area represents the identified set obtained by excluding a specific proxy, estimated with a fixed $ \tau=1.88 $. The gray area depicts the identified set under self-sign normalization only.
    \end{fignote}
    \label{fig:LOPO}
\end{figure}

\autoref{fig:LOPO} illustrates the resulting identified sets. The LOPO analysis demonstrates that the identified sets remain stable as long as the \cite{nakamura2018highfrequency} (NS) shock is included in the zoo.
Specifically, the identified sets for all other leave-one-out combinations (e.g., excluding GK or RR) cluster tightly around the benchmark full-zoo results.
In sharp contrast, excluding the NS proxy---and, to a lesser extent, the BRW proxy---causes the identified sets to expand dramatically, particularly for price responses.
This visual inspection provides a straightforward qualitative assessment of the sensitivity of the empirical claims to the composition of the proxy zoo.

This qualitative evidence is corroborated by the quantitative information measures reported in \autoref{tab:info_lopo}. Two patterns emerge.
First, the last column shows that the full zoo with eight proxies substantially tightens identification, reducing the widths of the identified sets by 86\% relative to the sign-only baseline. This marked improvement is expected, given that the baseline imposes only uninformative self-sign restrictions.
Second, excluding individual proxies reveals significant heterogeneity in information content. To quantify this, the second row reports the relative information $ \Delta_{\ell}(\tau^{\ast}) $. As is clear, the NS shock is by far the most informative proxy in the zoo: excluding it substantially reduces the information of the zoo by 21 percentage points. In contrast, removing the \cite{miranda-agrippino2021transmission} (MR) shock has no material effect on the identified set.\footnote{The information content of the zoo excluding MR ($0.87$) is slightly higher than the information of the full zoo ($0.86$). This discrepancy arises because the optimization problem guarantees only local optima.}

Overall, the LOPO exercise clarifies the critical assumptions driving the empirical findings.
If researchers believe that the NS proxy is significantly more contaminated than implied by $ \tau^{\ast} $---for example, due to new evidence of contamination---the remaining proxy zoo would lack the information necessary to identify contractionary effects of monetary policy.

\begin{table}[h!]
    \centering
    \caption{Proxy Information Analysis (LOPO)}
    \label{tab:info_lopo}
    \begin{threeparttable}
        \begin{tabular}{lccccccccc}
            \toprule
            Excluded Proxy                      & MR   & BS   & RR   & GSST & GK   & NS   & JK   & BRW  & None \\
            \midrule
            $ \kappa(\mathcal{M},\tau^{\ast}) $ & 0.87 & 0.82 & 0.82 & 0.82 & 0.78 & 0.65 & 0.85 & 0.78 & 0.86 \\
            $ \Delta_{\ell}(\tau^{\ast}) $      & 0.00 & 0.04 & 0.04 & 0.05 & 0.08 & 0.21 & 0.01 & 0.08 & 0.00 \\
            \bottomrule
        \end{tabular}
        \begin{tablenotes}[para, flushleft]
            \footnotesize
            \emph{Notes:} Each column reports (relative) information measures for the proxy zoo excluding the specified proxy, evaluated at the breakdown value $ \tau^{\ast}=1.88 $.
            Row $ \kappa(\mathcal{M},\tau^{\ast}) $ measures the information of the corresponding proxy zoo, and Row $ \Delta_{\ell}(\tau^{\ast}) $ measures the relative information of each excluded proxy, as in Equation~\eqref{eq:lopo_informativeness}.
            Higher values of $\Delta_{\ell}(\tau^{\ast})$ indicate that the excluded proxy contributes more to tightening identification. Information measures $ \kappa(\mathcal{M},\tau^{\ast}) $ for each variable and horizon are reported by \autoref{fig:LOPO_Informativeness} in the appendix.
        \end{tablenotes}
    \end{threeparttable}
\end{table}

\section{Conclusion}\label{sec:conclusion}

This paper develops a framework for robust identification with external proxies that relaxes exact proxy exogeneity and treats proxy contamination as a matter of degree, rather than a binary property. By combining generalized ranking restrictions (GRR) with auxiliary economic constraints, we characterize identified sets of impulse responses, derive upper bounds on proxy quality using falsifiable information, and propose diagnostic tools that make transparent how empirical conclusions depend on assumptions about proxy contamination and the composition of the proxy zoo.
We illustrate the robustness and transparency of the proposed framework using contaminated proxies in a simulated New Keynesian DSGE model, where the conventional proxy-SVAR approach induces substantial bias.
In the empirical analysis, we show how the zoo of monetary policy proxies can be reconciled with economic theory under substantially weaker assumptions than those required by conventional IV approaches.

Several directions for future research follow naturally.
First, while the present analysis takes the proxy zoo as given, the geometric structure underlying our results highlights the role of proxy complementarity in sharpening identification with external information. Proxies contribute to identification by imposing distinct, non-redundant restrictions on admissible structural representations, so that the resulting identified sets reflect the joint consistency of external information across the proxy zoo, rather than hinging on any single proxy in isolation.
Formalizing proxy construction as a search for complementarity among proxies, instead of a pursuit of perfect exogeneity, is an important avenue for future work.

Second, the partial identification of the proxy quality parameter raises the possibility of formal diagnostic procedures. In the empirical application, the fact that the upper bound on proxy quality is finite indicates that the joint restrictions implied by the proxy zoo are incompatible with exact exogeneity for all proxies.
Adapting this framework to develop flexible exogeneity tests is a natural extension.

Finally, although our analysis focuses on proxy-SVARs, the core insight that proxy complementarity within a zoo can deliver narrow and economically meaningful identified sets even under endogeneity is more general. Extending the framework to regression settings with mildly endogenous instruments, where the geometric structure is less explicit, may help clarify how information contained in the direction of endogeneity bias across multiple instruments can be exploited for identification.

Taken together, these results emphasize that the identifying content of external information is inherently joint, and that identification depends less on the validity of individual proxies than on the structure and interaction of the proxy zoo as a whole.

\afterpage{\clearpage}
\newpage
\begin{flushleft}
    \bibliographystyle{apalike}
    \bibliography{proxy_zoo}

@article{andrews2017measuring,
  title = {Measuring the Sensitivity of Parameter Estimates to Estimation Moments},
  author = {Andrews, Isaiah and Gentzkow, Matthew and Shapiro, Jesse M.},
  year = 2017,
  month = nov,
  journal = {The Quarterly Journal of Economics},
  volume = {132},
  pages = {1553--1592}
}

@article{andrews2020transparency,
  title = {Transparency in Structural Research},
  author = {Andrews, Isaiah and Gentzkow, Matthew and Shapiro, Jesse M.},
  year = 2020,
  month = oct,
  journal = {Journal of Business \& Economic Statistics},
  volume = {38},
  pages = {711--722},
  publisher = {Taylor \& Francis}
}

@article{angelini2019exogenous,
  title = {Exogenous uncertainty and the identification of structural vector autoregressions with external instruments},
  author = {Angelini, Giovanni and Fanelli, Luca},
  year = 2019,
  journal = {Journal of Applied Econometrics},
  volume = {34},
  pages = {951--971},
  copyright = {\copyright{} 2019 John Wiley \& Sons, Ltd.}
}

@article{angelini2025invalid,
  title = {Invalid Proxies and Volatility Changes},
  author = {Angelini, Giovanni and Fanelli, Luca and Neri, Luca},
  year = 2025,
  month = dec,
  journal = {Journal of Business \& Economic Statistics},
  publisher = {Taylor \& Francis},
  copyright = {\copyright{} 2025 American Statistical Association}
}

@misc{angelini2025test,
  title = {A Test of Exogeneity in Structural Vector Autoregressions and Local Projections with External instruments},
  author = {Angelini, Giovanni and Cavaliere, Giuseppe and Fanelli, Luca},
  year = 2025
}

@article{antolin-diaz2018narrative,
  title = {Narrative Sign Restrictions for SVARs},
  author = {{Antol{\'i}n-D{\'i}az}, Juan and {Rubio-Ram{\'i}rez}, Juan F.},
  year = 2018,
  month = oct,
  journal = {American Economic Review},
  volume = {108},
  pages = {2802--2829}
}

@article{arias2018inference,
  title = {Inference Based on Structural Vector Autoregressions Identified with Sign and Zero Restrictions: Theory and Applications},
  author = {Arias, Jonas E. and {Rubio-Ram{\'i}rez}, Juan F. and Waggoner, Daniel F.},
  year = 2018,
  journal = {Econometrica},
  volume = {86},
  eprint = {44955981},
  eprinttype = {jstor},
  pages = {685--720},
  publisher = {[Wiley, The Econometric Society]}
}

@misc{barnichon2025innovations,
  title = {Innovations Meet Narratives: Improving the Power-credibility Trade-off in Macro},
  author = {Barnichon, R{\'e}gis and Mesters, Geert},
  year = 2025,
  month = jan
}

@article{bauer2023reassessment,
  title = {A Reassessment of Monetary Policy Surprises and High-Frequency Identification},
  author = {Bauer, Michael D. and Swanson, Eric T.},
  year = 2023,
  month = may,
  journal = {NBER Macroeconomics Annual},
  volume = {37},
  pages = {87--155},
  publisher = {The University of Chicago Press}
}

@article{baumeister2015sign,
  title = {Sign Restrictions, Structural Vector Autoregressions, and Useful Prior Information},
  author = {Baumeister, Christiane and Hamilton, James D.},
  year = 2015,
  journal = {Econometrica},
  volume = {83},
  pages = {1963--1999}
}

@article{baumeister2019structural,
  title = {Structural Interpretation of Vector Autoregressions with Incomplete Identification: Revisiting the Role of Oil Supply and Demand Shocks},
  author = {Baumeister, Christiane and Hamilton, James D.},
  year = 2019,
  month = may,
  journal = {American Economic Review},
  volume = {109},
  pages = {1873--1910}
}

@article{beaudry2006stock,
  title = {Stock Prices, News, and Economic Fluctuations},
  author = {Beaudry, Paul and Portier, Franck},
  year = 2006,
  month = sep,
  journal = {American Economic Review},
  volume = {96},
  pages = {1293--1307}
}

@book{ben-tal2009robust,
  title = {Robust optimization},
  author = {{Ben-Tal}, Aharon and El Ghaoui, Laurent and Nemirovskij, Arkadij Semenovi{\v c}},
  year = 2009,
  publisher = {Princeton university press},
  address = {Princeton}
}

@article{benzeev2015investmentspecific,
  title = {Investment-Specific News Shocks and U.S. Business Cycles},
  author = {Ben Zeev, Nadav and Khan, Hashmat},
  year = 2015,
  journal = {Journal of Money, Credit and Banking},
  volume = {47},
  pages = {1443--1464},
  copyright = {\copyright{} 2015 The Ohio State University}
}

@article{benzeev2017chronicle,
  title = {Chronicle of a War Foretold: The Macroeconomic Effects of Anticipated Defence Spending Shocks},
  author = {Ben Zeev, Nadav and Pappa, Evi},
  year = 2017,
  journal = {The Economic Journal},
  volume = {127},
  pages = {1568--1597},
  copyright = {\copyright{} 2015 Royal Economic Society}
}

@article{benzeev2018what,
  title = {What can we learn about news shocks from the late 1990s and early 2000s boom-bust period?},
  author = {Ben Zeev, Nadav},
  year = 2018,
  month = feb,
  journal = {Journal of Economic Dynamics and Control},
  volume = {87},
  pages = {94--105}
}

@book{boyd2023convex,
  title = {Convex optimization},
  author = {Boyd, Stephen P. and Vandenberghe, Lieven},
  year = 2023,
  edition = {Version 29},
  publisher = {Cambridge University Press},
  address = {Cambridge New York Melbourne New Delhi Singapore}
}

@article{braun2023identification,
  title = {Identification of SVAR Models by Combining Sign Restrictions With External Instruments},
  author = {Braun, Robin and Br{\"u}ggemann, Ralf},
  year = 2023,
  month = oct,
  journal = {Journal of Business \& Economic Statistics},
  volume = {41},
  pages = {1077--1089},
  publisher = {ASA Website}
}

@article{brennan2025monetary,
  title = {Monetary Policy Shocks: Data or Methods?},
  author = {Brennan, Connor M. and Jacobson, Margaret M. and Matthes, Christian and Walker, Todd B.},
  year = 2025,
  month = jun,
  journal = {The B.E. Journal of Macroeconomics},
  volume = {25},
  pages = {595--659},
  publisher = {De Gruyter},
  copyright = {De Gruyter expressly reserves the right to use all content for commercial text and data mining within the meaning of Section 44b of the German Copyright Act.}
}

@article{bruns2024testing,
  title = {Testing for strong exogeneity in Proxy-VARs},
  author = {Bruns, Martin and Keweloh, Sascha A.},
  year = 2024,
  month = oct,
  journal = {Journal of Econometrics},
  volume = {245},
  pages = {105876}
}

@article{bu2021unified,
  title = {A Unified Measure of Fed Monetary Policy Shocks},
  author = {Bu, Chunya and Rogers, John and Wu, Wenbin},
  year = 2021,
  month = mar,
  journal = {Journal of Monetary Economics},
  volume = {118},
  pages = {331--349}
}

@article{carriero2024blended,
  title = {Blended identification in structural VARs},
  author = {Carriero, Andrea and Marcellino, Massimiliano and Tornese, Tommaso},
  year = 2024,
  month = apr,
  journal = {Journal of Monetary Economics},
  pages = {103581}
}

@article{cinelli2020making,
  title = {Making Sense of Sensitivity: Extending Omitted Variable Bias},
  author = {Cinelli, Carlos and Hazlett, Chad},
  year = 2020,
  month = feb,
  journal = {Journal of the Royal Statistical Society Series B: Statistical Methodology},
  volume = {82},
  pages = {39--67}
}

@article{conley2012plausibly,
  title = {Plausibly Exogenous},
  author = {Conley, Timothy G. and Hansen, Christian B. and Rossi, Peter E.},
  year = 2012,
  journal = {The Review of Economics and Statistics},
  volume = {94},
  eprint = {41349174},
  eprinttype = {jstor},
  pages = {260--272},
  publisher = {The MIT Press}
}

@misc{fernald2014quarterly,
  title = {A Quarterly, Utilization-Adjusted Series on Total Factor Productivity},
  author = {Fernald, John G.},
  year = 2014,
  month = apr,
  publisher = {Federal Reserve Bank of San Francisco Working Paper 2012-19},
  archiveprefix = {Federal Reserve Bank of San Francisco Working Paper 2012-19}
}

@article{fisher2010using,
  title = {Using Stock Returns to Identify Government Spending Shocks},
  author = {Fisher, Jonas D.M. and Peters, Ryan},
  year = 2010,
  month = may,
  journal = {The Economic Journal},
  volume = {120},
  pages = {414--436}
}

@article{francis2014flexible,
  title = {A Flexible Finite-Horizon Alternative to Long-Run Restrictions with an Application to Technology Shocks},
  author = {Francis, Neville and Owyang, Michael T. and Roush, Jennifer E. and DiCecio, Riccardo},
  year = 2014,
  month = oct,
  journal = {The Review of Economics and Statistics},
  volume = {96},
  pages = {638--647}
}

@article{gertler2015monetary,
  title = {Monetary Policy Surprises, Credit Costs, and Economic Activity},
  author = {Gertler, Mark and Karadi, Peter},
  year = 2015,
  month = jan,
  journal = {American Economic Journal: Macroeconomics},
  volume = {7},
  pages = {44--76}
}

@article{gurkaynak2005actions,
  title = {Do Actions Speak Louder Than Words? The Response of Asset Prices to Monetary Policy Actions and Statements},
  author = {G{\"u}rkaynak, Refet S. and Sack, Brian and Swanson, Eric},
  year = 2005,
  month = may,
  journal = {International Journal of Central Banking},
  volume = {1},
  pages = {55--93}
}

@article{hansen1982large,
  title = {Large Sample Properties of Generalized Method of Moments Estimators},
  author = {Hansen, Lars Peter},
  year = 1982,
  month = jul,
  journal = {Econometrica},
  volume = {50},
  eprint = {1912775},
  eprinttype = {jstor},
  pages = {1029}
}

@article{jarocinski2020deconstructing,
  title = {Deconstructing Monetary Policy Surprises--- The Role of Information Shocks},
  author = {Jaroci{\'n}ski, Marek and Karadi, Peter},
  year = 2020,
  month = apr,
  journal = {American Economic Journal: Macroeconomics},
  volume = {12},
  pages = {1--43}
}

@article{jentsch2022asymptotically,
  title = {Asymptotically Valid Bootstrap Inference for Proxy SVARs},
  author = {Jentsch, Carsten and Lunsford, Kurt G.},
  year = 2022,
  journal = {Journal of Business and Economic Statistics},
  volume = {40},
  pages = {1876--1891}
}

@article{kanzig2021macroeconomic,
  title = {The Macroeconomic Effects of Oil Supply News: Evidence from OPEC Announcements},
  author = {K{\"a}nzig, Diego R.},
  year = 2021,
  month = apr,
  journal = {American Economic Review},
  volume = {111},
  pages = {1092--1125}
}

@article{keweloh2025estimating,
  title = {Estimating Fiscal Multipliers by Combining Statistical Identification with Potentially Endogenous Proxies},
  author = {Keweloh, Sascha A and Klein, Mathias and Pr{\"u}ser, Jan},
  year = 2025,
  month = nov,
  journal = {The Econometrics Journal},
  pages = {utaf027}
}

@book{kilian2017structural,
  title = {Structural Vector Autoregressive Analysis},
  author = {Kilian, Lutz and L{\"u}tkepohl, Helmut},
  year = 2017,
  month = nov,
  publisher = {Cambridge University Press}
}

@article{kiviet2020testing,
  title = {Testing the Impossible: Identifying Exclusion Restrictions},
  author = {Kiviet, Jan F.},
  year = 2020,
  month = oct,
  journal = {Journal of Econometrics},
  volume = {218},
  pages = {294--316}
}

@article{leeper2012quantitative,
  title = {Quantitative Effects of Fiscal Foresight},
  author = {Leeper, Eric M and Richter, Alexander W and Walker, Todd B},
  year = 2012,
  month = may,
  journal = {American Economic Journal: Economic Policy},
  volume = {4},
  pages = {115--144}
}

@article{ludvigson2021uncertainty,
  title = {Uncertainty and Business Cycles: Exogenous Impulse or Endogenous Response?},
  author = {Ludvigson, Sydney C. and Ma, Sai and Ng, Serena},
  year = 2021,
  month = oct,
  journal = {American Economic Journal: Macroeconomics},
  volume = {13},
  pages = {369--410}
}

@book{manski2003partial,
  title = {Partial Identification of Probability Distributions},
  author = {Manski, Charles F.},
  year = 2003,
  publisher = {Springer},
  address = {New York}
}

@article{masten2020inference,
  title = {Inference on breakdown frontiers},
  author = {Masten, Matthew A. and Poirier, Alexandre},
  year = 2020,
  journal = {Quantitative Economics},
  volume = {11},
  pages = {41--111},
  copyright = {http://creativecommons.org/licenses/by-nc/4.0/}
}

@article{masten2021salvaging,
  title = {Salvaging Falsified Instrumental Variable Models},
  author = {Masten, Matthew A. and Poirier, Alexandre},
  year = 2021,
  journal = {Econometrica},
  volume = {89},
  pages = {1449--1469}
}

@article{mertens2013dynamic,
  title = {The Dynamic Effects of Personal and Corporate Income Tax Changes in the United States},
  author = {Mertens, Karel and Ravn, Morten O},
  year = 2013,
  month = jun,
  journal = {American Economic Review},
  volume = {103},
  pages = {1212--1247}
}

@article{miranda-agrippino2021transmission,
  title = {The Transmission of Monetary Policy Shocks},
  author = {{Miranda-Agrippino}, Silvia and Ricco, Giovanni},
  year = 2021,
  journal = {American Economic Journal: Macroeconomics},
  volume = {13},
  pages = {74--107}
}

@article{nakamura2018highfrequency,
  title = {High-Frequency Identification of Monetary Non-Neutrality: The Information Effect},
  author = {Nakamura, Emi and Steinsson, J{\'o}n},
  year = 2018,
  month = aug,
  journal = {Quarterly Journal of Economics},
  volume = {133},
  pages = {1283--1330}
}

@article{nguyen2025bayesian,
  title = {Bayesian inference in proxy SVARs with incomplete identification: Re-evaluating the validity of monetary policy instruments},
  author = {Nguyen, Lam},
  year = 2025,
  month = jul,
  journal = {Journal of Monetary Economics},
  pages = {103813}
}

@article{ottonello2025financial,
  title = {Financial Intermediaries and the Macroeconomy: Evidence from a High-Frequency Identification},
  author = {Ottonello, Pablo and Song, Wenting},
  year = 2025,
  month = nov,
  journal = {The Economic Journal},
  pages = {ueaf119},
  copyright = {https://academic.oup.com/journals/pages/open\_access/funder\_policies/chorus/standard\_publication\_model}
}

@article{plagborg-moller2021local,
  title = {Local Projections and VARs Estimate the Same Impulse Responses},
  author = {{Plagborg-M{\o}ller}, Mikkel and Wolf, Christian K.},
  year = 2021,
  journal = {Econometrica},
  volume = {89},
  pages = {955--980}
}

@incollection{ramey2016macroeconomic,
  title = {Macroeconomic Shocks and Their Propagation},
  booktitle = {Handbook of Macroeconomics},
  author = {Ramey, V.A.},
  year = 2016,
  volume = {2},
  pages = {71--162},
  publisher = {Elsevier},
  copyright = {https://www.elsevier.com/tdm/userlicense/1.0/}
}

@article{ramey2018government,
  title = {Government Spending Multipliers in Good Times and in Bad: Evidence from US Historical Data},
  author = {Ramey, Valerie A and Zubairy, Sarah},
  year = 2018,
  month = mar,
  journal = {Journal of Political Economy},
  volume = {126},
  pages = {850--901}
}

@article{romer2004new,
  title = {A New Measure of Monetary Shocks: Derivation and Implications},
  author = {Romer, Christina D and Romer, David H},
  year = 2004,
  month = sep,
  journal = {American Economic Review},
  volume = {94},
  pages = {1055--1084}
}

@article{sargan1958estimation,
  title = {The Estimation of Economic Relationships using Instrumental Variables},
  author = {Sargan, J. D.},
  year = 1958,
  month = jul,
  journal = {Econometrica},
  volume = {26},
  eprint = {1907619},
  eprinttype = {jstor},
  pages = {393}
}

@article{schlaak2023monetary,
  title = {Monetary policy, external instruments, and heteroskedasticity},
  author = {Schlaak, Thore and Rieth, Malte and Podstawski, Maximilian},
  year = 2023,
  journal = {Quantitative Economics},
  volume = {14},
  pages = {161--200},
  copyright = {Copyright \copyright{} 2023 The Authors.}
}

@article{smets2007shocks,
  title = {Shocks and Frictions in US Business Cycles: A Bayesian DSGE Approach},
  author = {Smets, Frank and Wouters, Rafael},
  year = 2007,
  month = jun,
  journal = {American Economic Review},
  volume = {97},
  pages = {586--606}
}

@article{stock2018identification,
  title = {Identification and Estimation of Dynamic Causal Effects in Macroeconomics Using External Instruments},
  author = {Stock, James H. and Watson, Mark W.},
  year = 2018,
  month = may,
  journal = {The Economic Journal},
  volume = {128},
  pages = {917--948}
}

@article{uhlig2005what,
  title = {What are the effects of monetary policy on output? Results from an agnostic identification procedure},
  author = {Uhlig, Harald},
  year = 2005,
  month = mar,
  journal = {Journal of Monetary Economics},
  volume = {52},
  pages = {381--419}
}

@article{vanderweele2017sensitivity,
  title = {Sensitivity Analysis in Observational Research: Introducing the E-Value},
  author = {VanderWeele, Tyler J. and Ding, Peng},
  year = 2017,
  month = aug,
  journal = {Annals of Internal Medicine},
  volume = {167},
  pages = {268--274},
  publisher = {American College of Physicians}
}
\end{flushleft}

\newpage
\begin{appendices}
    
\setcounter{figure}{0}
\setcounter{table}{0}
\setcounter{equation}{0}

\renewcommand{\thefigure}{\thesection.\arabic{figure}}
\renewcommand{\thetable}{\thesection.\arabic{table}}
\renewcommand{\theequation}{\thesection.\arabic{equation}}
    \section{Computational details}\label{app:computation}
    The objective function is linear in $O_{\bullet 1}$, but the constraint set is nonconvex due to the orthogonality requirement $O \in \mathcal{O}(n)$. To handle this constraint, we parameterize the space of rotation matrices via the matrix exponential. Any orthogonal matrix $O$ with $\det(O) = 1$ can be written as $O = \exp(S)$ where $S$ is skew-symmetric, i.e., $S' = -S$.\footnote{For any $(n\times n)$ skew-symmetric $S$, we have $\exp(S) \exp(S)' = \exp(S)\exp(-S) = I_{n} = OO'$.} This parameterization maps the $n(n-1)/2$ free parameters in $S$---the strictly lower-triangular entries---to $\mathcal{O}(n)$, converting the orthogonality-constrained problem into an unconstrained one over $\mathbb{R}^{n(n-1)/2}$ \citep[cf.][]{arias2018inference}. The absolute value in the ranking restrictions is handled by \textsf{fmincon} in MATLAB.

    In general, it is difficult to guarantee a global optimum of the optimization problem \eqref{eq:optim_obj}. To mitigate the risk of converging to local optima, we employ warm starts and solve the optimization problem sequentially over horizons $h$. That is, we use the optimizer $O^{(h)}$ obtained in horizon $h$ as the initial value for optimization at horizon $h+1$. For horizon $h=0$, we simply use the short-run restrictions as the initial point, which corresponds to $O^{(0)}=I_n$.

    \subsection{Solving for the upper bound \texorpdfstring{$ \bar{\tau}$}{tau}}
    We now show how to transform the computation of the upper bound $ \bar{\tau} $ in \eqref{eq:tau_bound_tight} into a tractable optimization problem.
    Note that $ \cot (\cdot) $ is strictly decreasing on $ (0,\pi) $. We have
    
 \begin{align*}
\cot\!\left( \min_{q \in \mathcal{G}_{\textup{sign}}} \max_{\ell=1,\ldots,k} \alpha_{\ell,q} \right)
&= \max_{q \in \mathcal{G}_{\textup{sign}}} 
   \cot\!\left( \max_{\ell=1,\ldots,k} \alpha_{\ell,q} \right) \\[0.5em]
&= \max_{q \in \mathcal{G}_{\textup{sign}}} 
   \min_{\ell=1,\ldots,k} \cot(\alpha_{\ell,q}) \\[0.5em]
&= \max_{q \in \mathcal{G}_{\textup{sign}}} 
   \min_{\ell=1,\ldots,k}
   \frac{\cos(\alpha_{\ell,q})}{\sqrt{1 - \cos^2(\alpha_{\ell,q})}} \\[0.5em]
&= \max_{q \in \mathcal{G}_{\textup{sign}}} 
   \min_{\ell=1,\ldots,k}
   \frac{\frac{M_\ell' q}{\|M_\ell\|\,\|q\|}}
        {\sqrt{1 - \left(\frac{M_\ell' q}{\|M_\ell\|\,\|q\|}\right)^2}} \\[0.5em]
&= \max_{q \in \mathcal{G}_{\textup{sign}}} 
   \min_{\ell=1,\ldots,k}
   \frac{M_\ell' q}{\sqrt{\|M_\ell\|^2 - (M_\ell' q)^2}} \, .
\end{align*}
    In the second last equality we have used the definition of the angle $\alpha_{\ell,q}$:
    \begin{equation*}
        \cos \alpha_{\ell,q} =  \frac{M_{\ell}' q}{\|M_{\ell}\| \|q\|} ~.
    \end{equation*}
    And the last equality follows from the fact that $\|q\| = 1$ for all $q \in \mathcal{G}_{\textup{sign}}$. Therefore, the upper bound $ \bar{\tau} $ can be computed by solving the following optimization problem:
    \begin{equation*}
        \bar{\tau} = \sqrt{n-1}\max_{q \in \mathcal{G}_{\textup{sign}}} \min_{\ell=1,\ldots,k }
        \frac{M_{\ell}' q}{\sqrt{\|M_{\ell}\|^{2} - (M_{\ell}' q)^{2}}}~, \qquad ||q||=1~.
    \end{equation*}

    The above problem is a classical max-min problem in the robust optimization literature \citep[see, e.g.,][]{ben-tal2009robust}. We convert it into a standard constrained optimization problem. Specifically, we can view the inner minimized value $ \frac{M_{\ell}' q}{\sqrt{\|M_{\ell}\|^{2} - (M_{\ell}' q)^{2}}} $ as a function of the column of interest $ q $:
    \begin{equation}\label{eq:fq}
        f(q) = \min_{\ell=1,\ldots,k }
        \frac{M_{\ell}' q}{\sqrt{\|M_{\ell}\|^{2} - (M_{\ell}' q)^{2}}}~.
    \end{equation}
    The optimization problem reduces to maximizing $ f(q) $ over $ q \in \mathcal{G}_{\textup{sign}} $, which, without loss of generality, can be expressed as a set of inequality constraints $ g_{r}(q) \ge  0 $ for $  r=1,\ldots,R $ with $ R $ the total number of constraints on $ q $. That is, we aim to solve
    \begin{equation}\label{eq:maxminproblem}
        \begin{aligned}
            \max_{q \in \mathbb{R}^{n}} & \quad f(q)                               \\
            \text{s.t.}                 & \quad g_{r}(q) \ge  0,\quad r=1,\ldots,R \\
                                        & \quad \| q \| = 1                        \\
        \end{aligned}
    \end{equation}
    We reformulate the max-min problem using a standard epigraph construction with an auxiliary variable, following the treatment of pointwise minima in \citet{boyd2023convex}.
    We introduce the auxiliary variable $ v $ and suppose it satisfies $ v\le \frac{M_{\ell}' q}{\sqrt{\|M_{\ell}\|^{2} - (M_{\ell}' q)^{2}}} $ for all $ \ell=1,\ldots,k $. By construction, we have \textit{for all feasible pair} $ q $ and $ v $:
    \begin{align*}
        v & \le \frac{M_{\ell}' q}{\sqrt{\|M_{\ell}\|^{2} - (M_{\ell}' q)^{2}}} \quad\forall ~\ell=1,\ldots,k \\
        v & \le \min_{\ell=1,\ldots,k }
        \frac{M_{\ell}' q}{\sqrt{\|M_{\ell}\|^{2} - (M_{\ell}' q)^{2}}}=f(q)~.
    \end{align*}
    For any fixed $ q $, the largest feasible $ v $ equals $ f(q) $. Therefore, if we maximize over both $ q $ and $ v $ subject to the above constraints, the optimal $ v^{\star} $ must equal $ \max_{q}f(q) $.

    So, problem~\eqref{eq:maxminproblem} can be written in the epigraph form
    \begin{equation}\label{eq:tau_ratio_epigraph}
        \begin{aligned}
            \max_{q,v}  & \quad v                                                                                                    \\
            \text{s.t.} & \quad v \leq \frac{M_{\ell}' q}{\sqrt{\|M_{\ell}\|^{2} - (M_{\ell}' q)^{2}}} \quad\forall ~\ell=1,\ldots,k \\
                        & \quad g_{r}(q) \ge  0,\quad r=1,\ldots,R                                                                   \\
                        & \quad \| q \| = 1                                                                                          \\
        \end{aligned}
    \end{equation}

    \paragraph{Equivalent epigraph formulation and numerical stability.}
    Consider the optimization problem~\eqref{eq:maxminproblem}.
    For any feasible $q$, define
    \begin{equation}\label{eq:c_def}
        c_{\ell}(q) := \frac{M_{\ell}' q}{\|M_{\ell}\|} \in [0,1].
    \end{equation}
    Since $M_{\ell}' q = c_{\ell}(q)\,\|M_{\ell}\|$, substituting into the objective of~\eqref{eq:maxminproblem} yields
    \begin{equation}\label{eq:fq_ratio}
        f(q)
        =
        \min_{\ell}
        \frac{c_{\ell}(q)}{\sqrt{1-c_{\ell}(q)^2}}.
    \end{equation}
    Accordingly, the first constraint in \eqref{eq:tau_ratio_epigraph} can be written as
    \begin{equation}\label{eq:g_def}
        v \le h\!\left(c_{\ell}(q)\right),
        \qquad
        h(c) := \frac{c}{\sqrt{1-c^{2}}}.
    \end{equation}
    The function $h(\cdot)$ is strictly increasing on $(-1,1)$.
    Therefore, for any fixed $q$ satisfying the constraints,
    \begin{equation}\label{eq:min_g_identity}
        \min_{\ell} h\!\left(c_{\ell}(q)\right)
        =
        h\!\left(\min_{\ell} c_{\ell}(q)\right).
    \end{equation}
    It follows that the optimal value of \eqref{eq:maxminproblem} equals
    \begin{equation}\label{eq:g_of_cstar}
        h(c^{\star}),
        \qquad
        c^{\star}
        :=
        \max_{q \in \mathcal{G}_{\textup{sign}},\,\|q\|=1}
        \min_{\ell=1,\ldots,k}
        c_{\ell}(q).
    \end{equation}

    Although problem~\eqref{eq:tau_ratio_epigraph} is mathematically correct, directly enforcing
    the ratio constraints in \eqref{eq:g_def} is numerically unstable. The mapping
    $h(c)=c/\sqrt{1-c^2}$ has an unbounded derivative as $c\to 1$, which leads to ill-conditioned
    constraint Jacobians when the optimizer explores regions where $M_{\ell}' q$ is close to
    $\|M_{\ell}\|$.

    In computation, we therefore solve for $c^{\star}$ directly using the equivalent epigraph formulation
    \begin{equation}\label{eq:cosine_epigraph}
        \begin{aligned}
            \max_{q,t} \quad & t                                                \\
            \text{s.t.}\quad
                             & t \le c_{\ell}(q) \quad \forall \ell=1,\ldots,k, \\
                             & g_r(q) \ge 0, \quad r=1,\ldots,R,                \\
                             & \|q\| = 1 ,
        \end{aligned}
    \end{equation}
    which involves only bounded and smooth constraints.
    The upper bound in Theorem~\ref{thm:tau_identification} is then recovered as
    \begin{equation}\label{eq:tau_from_cstar}
        \bar{\tau}
        =
        \sqrt{n-1}\,
        \frac{c^{\star}}{\sqrt{1-(c^{\star})^{2}}}.
    \end{equation}
    This reformulation avoids placing the singular cotangent transformation inside the
    optimization problem while yielding exactly the same value of $\bar{\tau}$. Replacing consistently estimable quantities, we can solve the sample analog to obtain an estimate of $ \bar{\tau} $, using off-the-shelf constrained optimizers such as \textsf{fmincon} in MATLAB.

    \newpage
    \section{Geometric illustration}\label{app:geometry}

\begin{figure}[ht]
        \centering
        \begin{subfigure}[b]{0.45\textwidth}
            \includegraphics[width=\linewidth]{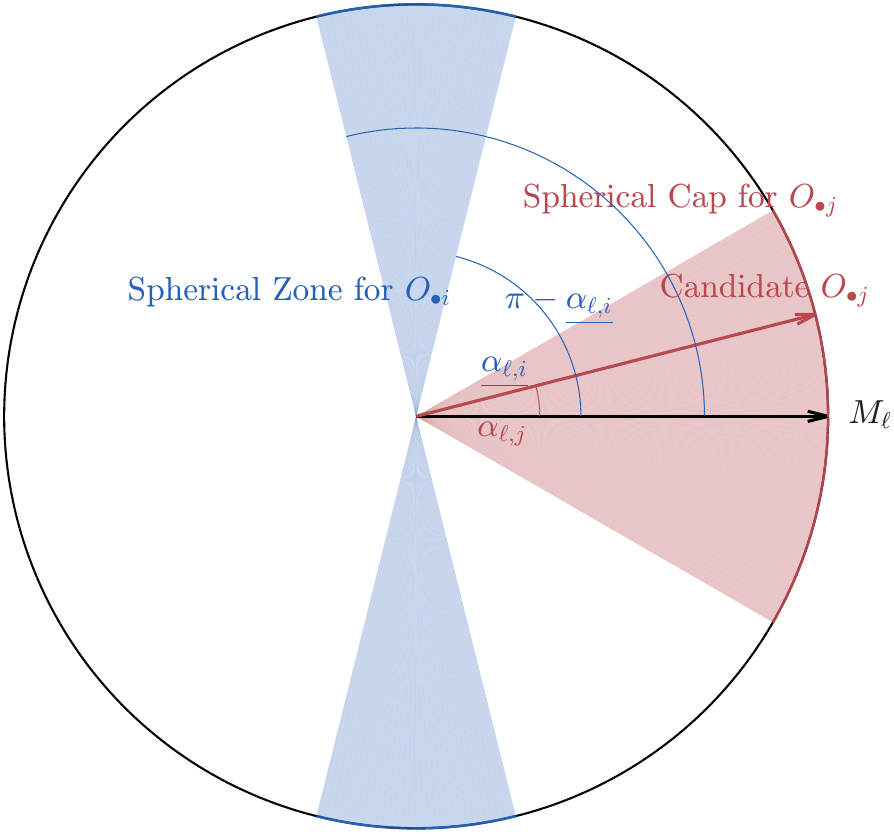}
            \caption{Spherical Cap Constraints}
            \label{fig:interpretation_single_cap}
        \end{subfigure}
        \begin{subfigure}[b]{0.45\textwidth}
            \includegraphics[width=\linewidth]{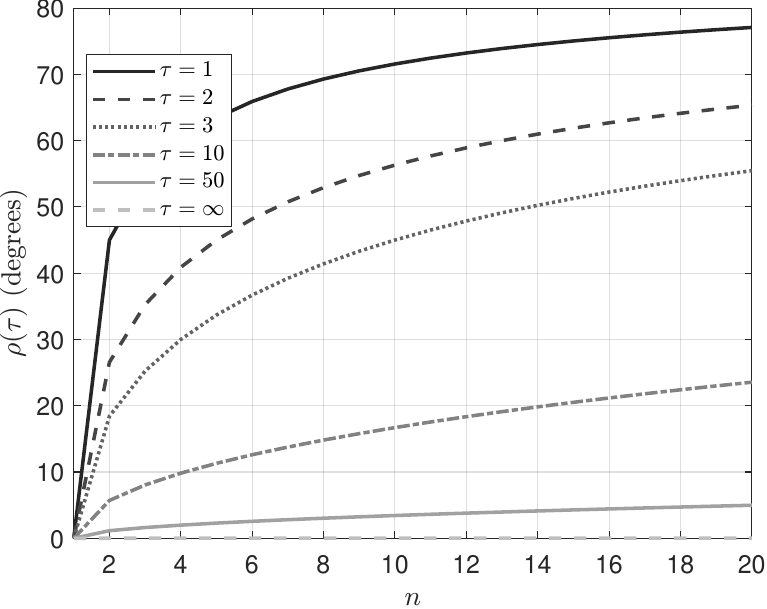}
            \caption{Upper Bound on Caps}
            \label{fig:rhotau}
        \end{subfigure}
        \caption{Geometric Interpretation of GRR.}
        \label{fig:interpretation_single_proxy}
    \end{figure}

    Lemma~\ref{lem:angular_bounds} shows how the ranking restrictions in \eqref{eq:rank_O_homog} translate into geometric constraints on the rotation matrix $O$. For each proxy $\ell$, the target column of the admissible rotations, $O_{\bullet 1}$, must lie within a spherical cap of angular radius $\rho(\tau)$ centered at the proxy-covariance direction $M_\ell$. Identification therefore reduces to restricting the location of $O_{\bullet 1}$ on the unit sphere.
    \autoref{fig:interpretation_single_cap} illustrates the constraints on the column $O_{\bullet 1}$ in two dimensions ($n=2$).\footnote{See Appendix~\ref{sec:framework} for formal definitions.}

    Two parameters govern the size of this cap.
    First, the dimension of the VAR affects the scope for contamination: a higher-dimensional system provides more non-target channels through which the proxy may correlate with other structural shocks. The penalty factor $\sqrt{n-1}$ reflects a conservative worst-case scenario in which contamination is evenly distributed across all $n-1$ non-target directions.
    Second, the proxy-quality parameter $\tau$ controls the tightness of the restriction. Larger values of $\tau$ correspond to higher-quality proxies that deliver more precise directional information about $O_{\bullet 1}$. In the limit as $\tau \to \infty$, the valid-IV assumption is recovered: the cap radius $\rho(\tau)$ collapses to zero and $O_{\bullet 1}$ aligns exactly with $M_\ell$.
    \autoref{fig:rhotau} illustrates the relationship between the VAR dimension $n$, proxy quality $\tau$, and the implied angular bound $\rho(\tau)$; see Lemma~\ref{lem:angular_bounds} for the formal bound.

    \paragraph{Geometric interpretation of $\bar{\tau}$ in Theorem~\ref{thm:tau_identification}}

\begin{figure}[htb!]
        \centering
        \includegraphics[width=0.5\linewidth]{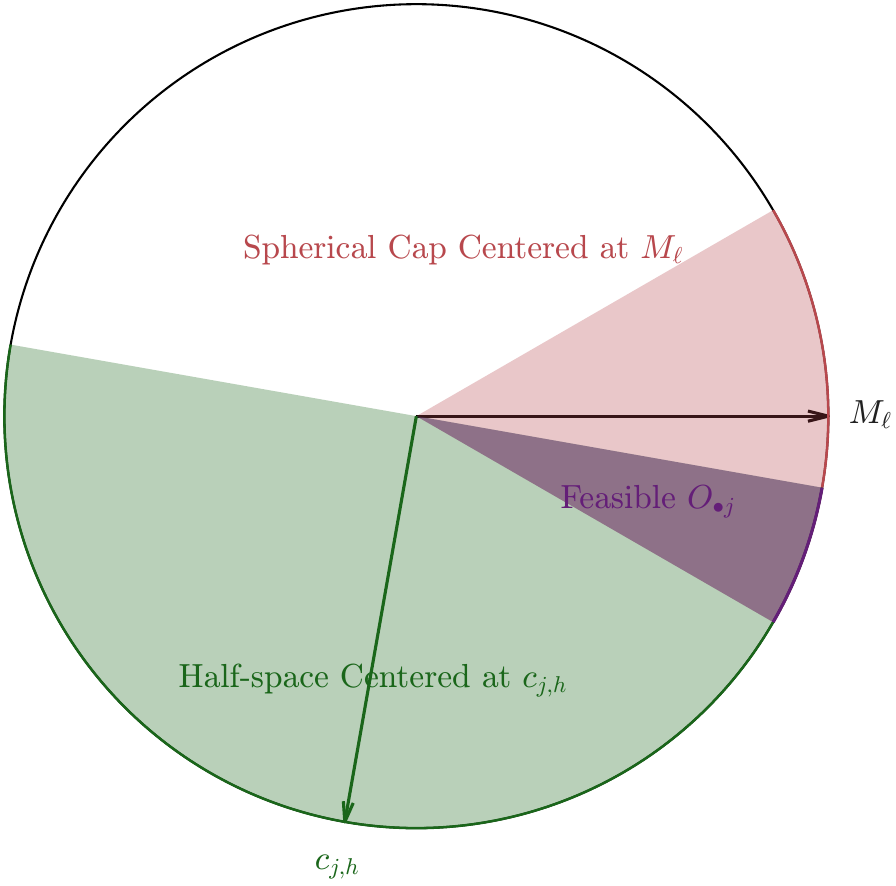}
        \caption{Complementarity Between Sign and GRR.}
        \label{fig:tau_identification_sign}
    \end{figure}

\begin{figure}[htb!]
        \centering
        \begin{tabular}{cc}
            \includegraphics[width=0.3\linewidth]{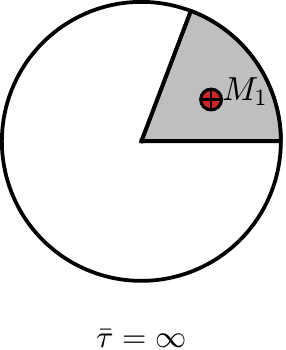}   & \hfill\hfill
            \includegraphics[width=0.3\linewidth]{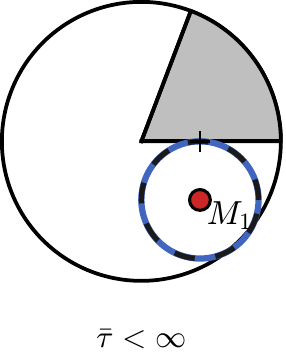}                 \\
            (a) One proxy inside                                               &
            (b) One proxy outside                                                             \\
            \includegraphics[width=0.3\linewidth]{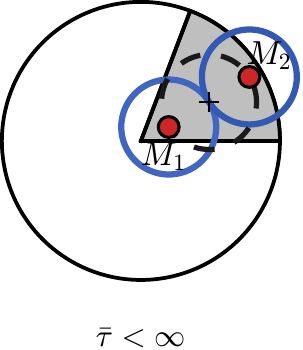} & \hfill\hfill
            \includegraphics[width=0.3\linewidth]{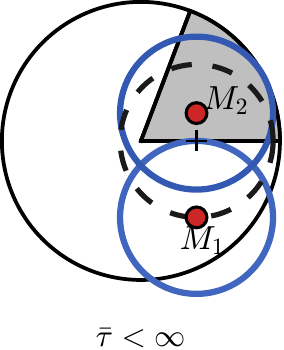}                     \\
            (c) Two proxies inside                                             &
            (d) One proxy inside, one outside
        \end{tabular}

        \caption{Geometric interpretation of $\bar{\tau}$ in Theorem~\ref{thm:tau_identification}.}
        \label{fig:tau_bar}
        \begin{fignote}[\linewidth]
            \emph{Notes:}
            Panel (a) shows the case of a single proxy whose direction lies inside the sign-restricted set; no finite upper bound on $\tau$ is implied and $\bar{\tau}=\infty$.
            Panel (b) shows a single proxy lying outside the sign-restricted set; the smallest cap intersecting the sign-restricted region yields a finite bound $\bar{\tau}<\infty$.
            Panel (c) shows two proxies inside the sign-restricted set; at $\tau=\bar{\tau}$ the minimum enclosing cap is unconstrained.
            Panel (d) shows one proxy inside and one outside the sign-restricted set; the bound $\bar{\tau}>1$ is determined by a \emph{constrained} minimum enclosing cap (dashed black circle), as characterized in Eq.~\eqref{eq:tau_bound_tight}.
        \end{fignote}
    \end{figure}

    Figures~\ref{fig:tau_identification_sign} and \ref{fig:tau_bar} provide geometric illustrations of the bound $\bar{\tau}$ characterized in Theorem~\ref{thm:tau_identification}. The figures illustrate two sources of identifying power.\footnote{The discussion here parallels the analytical characterization in Section~\ref{sec:bounds_tau}.}

    First, the identified set for the structural column $O_{\bullet 1}$ is obtained by intersecting the spherical caps implied by the proxies with the cones formed by the sign restrictions.
    \autoref{fig:tau_identification_sign} illustrates the case with a single proxy and a single sign restriction when $ n=2 $ (so the sign restriction corresponds to a half-space). As argued above, a larger proxy quality $\tau$ shrinks the spherical cap around the proxy moment vector $M_{\ell}$, and thus its intersection with the sign-restricted region. As long as the proxy does not lie \emph{inside} the sign cone---meaning that the proxy provides additional, conflicting information relative to the sign restrictions---we can identify a finite upper bound on the proxy-quality parameter that remains compatible with those sign restrictions.

    Second, \autoref{fig:tau_bar} depicts an aerial (equatorial) cross-section of the unit sphere, which suffices to visualize the relevant geometry. The gray region represents the set of directions admissible under auxiliary sign restrictions. Red points $(M_1, M_2)$ denote normalized proxy-covariance directions. Blue circles correspond to Euclidean cross-sections of the spherical caps centered at each proxy and indexed by $\tau$. The black dashed circle denotes the minimum enclosing cap---possibly constrained by the sign restrictions---that determines $\bar{\tau}$, as formalized in Eq.~\eqref{eq:tau_bound_tight}.

With a single proxy ($k=1$), the bound in \eqref{eq:tau_bound_tight} depends entirely on the smallest angular distance between
the proxy covariance vector and the set of admissible rotations defined by the sign restrictions.
When this distance is zero, so that 
$\min_{q \in \mathcal{G}_{\textup{sign}}} \alpha_{\ell,q} = 0,$ the proxy lies inside the sign-feasible cone and does not restrict $\tau_0$ from above (\autoref{fig:tau_bar}, Panel~(a)).
Conversely, a strictly positive distance implies that the proxy favors directions outside the sign-restricted set.
In this case, consistency between the proxy information and the sign restrictions requires allowing for contamination, which yields a finite upper bound $\bar{\tau} < \infty$
(\autoref{fig:tau_bar}, Panel~(b)).

With multiple proxies ($k \ge 2$), the bound additionally reflects disagreement across proxies.
Even when each proxy is individually compatible with the sign restrictions, the worst-case
misalignment $\max_{\ell=1,\ldots,k} \alpha_{\ell,q}$
for a given $q$ captures the possibility that proxies favor different directions within the sign-feasible cone.
Such conflicting information cannot be reconciled without allowing contamination in at least some proxies, thereby inducing a finite upper bound on $\tau_0$
(\autoref{fig:tau_bar}, Panel~(c)), in a manner analogous to overidentification in
classical IV settings \citep[e.g.,][]{sargan1958estimation}. 
Finally,  \autoref{fig:tau_bar}, Panel~(d) illustrates the case in which one proxy is incompatible with the sign restrictions. The bound $\bar{\tau}$ is obtained by computing the minimum enclosing cap of the proxy covariance directions on the sphere, constrained to have its center within the sign-feasible set (cf.\ Equation~\eqref{eq:tau_bound_tight}).

    \clearpage
    \section{Additional empirical results}

    \begin{figure}[ht]
        \centering
        \includegraphics[width=1\linewidth]{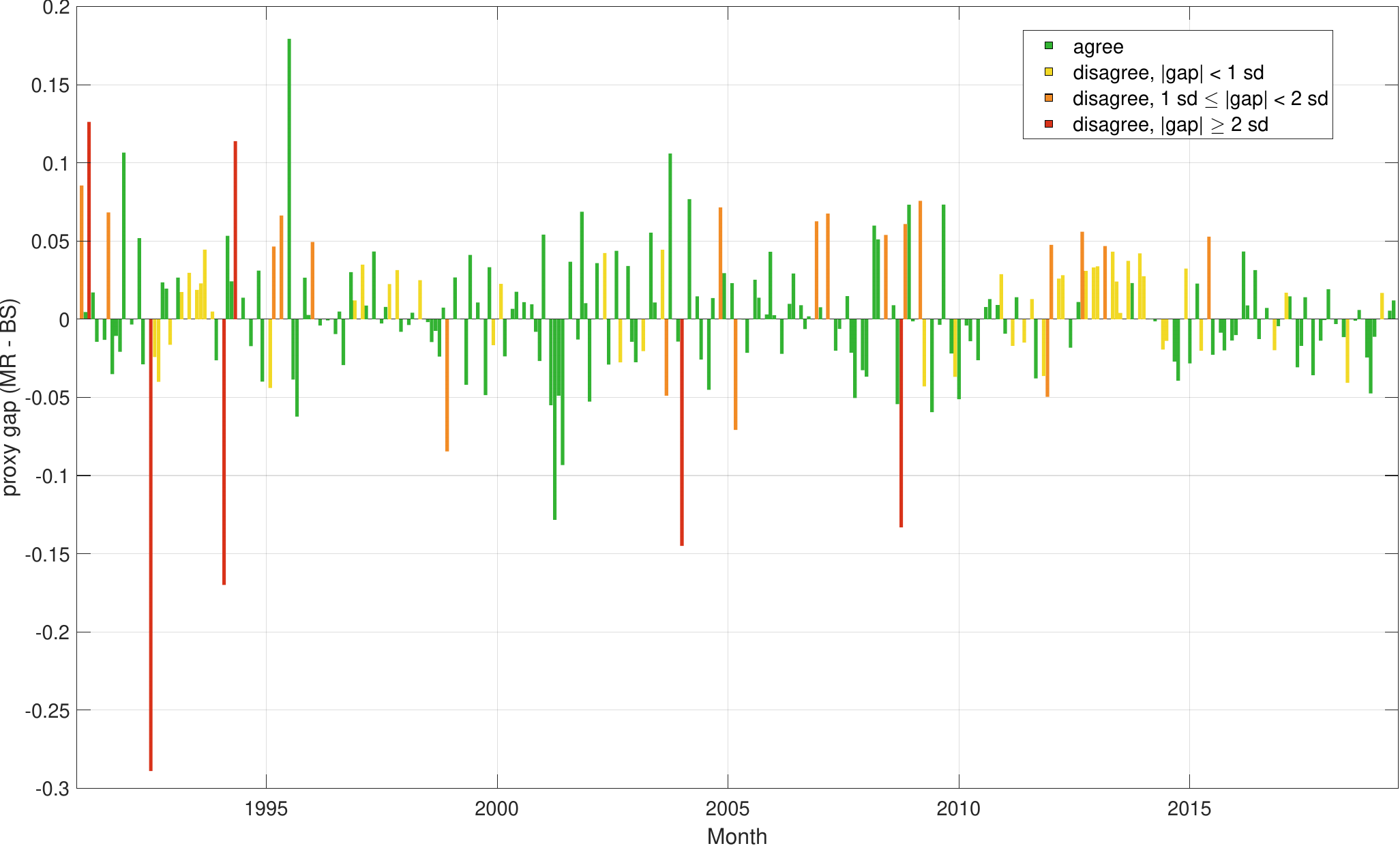}
        \caption{Sign Disagreement Between MR and BS Proxies}
        \begin{fignote}[\linewidth]
            \emph{Notes:} This figure illustrates the extent of sign disagreement between the \cite{miranda-agrippino2021transmission} (MR) proxy and the \cite{bauer2023reassessment} (BS) proxy. Darker colors indicate a larger discrepancy.
        \end{fignote}
        \label{fig:EMP-MRBS-Sign-Disagreement}
    \end{figure}

    \begin{figure}[ht]
        \centering
        \includegraphics[width=1\linewidth]{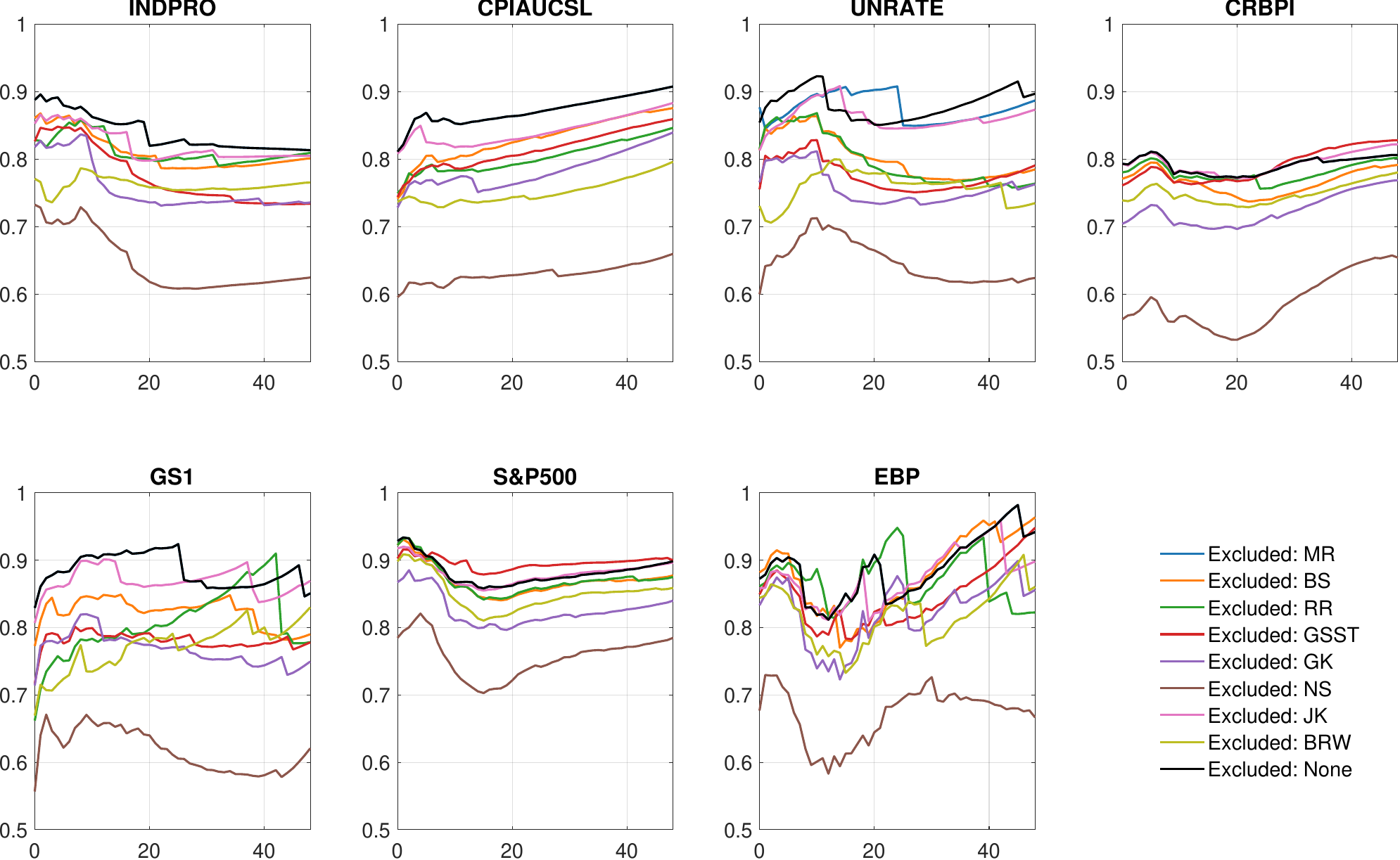}
        \caption{Proxy Zoo Identification: Information Measures}
        \begin{fignote}[\linewidth]
            \emph{Notes:} Information measure $ \kappa_{i,h}(\mathcal{M},\tau^{\ast}) $ of the monetary policy proxy zoo, computed at the breakdown value $ \tau^{\ast}=1.88 $. Each line represents the information with a particular excluded proxy from the full zoo. Higher values indicate that the proxy zoo contributes more to tightening identification relative to the self-sign normalization. See Definition~\ref{def:zoo_information}.
        \end{fignote}
        \label{fig:LOPO_Informativeness}
    \end{figure}

    \clearpage
    \newpage
    \begin{landscape}
        \begin{table}[ht]
            \centering
            \caption{Pairwise Correlations between Monetary Policy Proxies and Other Structural Shocks}
            \label{tab:pairwise_corr}
            \resizebox{1.4\textwidth}{!}{%
                \begin{threeparttable}
                    \begin{tabular}{lccccccccccccccccccc}
                        \toprule
                             & \multicolumn{3}{c}{\textbf{Oil Supply}} & \multicolumn{3}{c}{\textbf{Financial}} & \multicolumn{4}{c}{\textbf{Fiscal}} & \multicolumn{9}{c}{\textbf{Technology (TFP \& IST)}}                                                                                                                                                                                                                                                                                                                                                                                         \\
                        \cmidrule(lr){2-4} \cmidrule(lr){5-7} \cmidrule(lr){8-11}  \cmidrule(lr){12-20}
                             & \rotatebox{90}{K21\_S}                  & \rotatebox{90}{K21\_N}                 & \rotatebox{90}{BH19}                & \rotatebox{90}{OS25}                                 & \rotatebox{90}{OS25\_out} & \rotatebox{90}{OS25\_SR} & \rotatebox{90}{RZ18} & \rotatebox{90}{FP10} & \rotatebox{90}{BZP17} & \rotatebox{90}{LRW12} & \rotatebox{90}{BP06\_SR} & \rotatebox{90}{BP06\_LR} & \rotatebox{90}{F14} & \rotatebox{90}{F14\_U} & \rotatebox{90}{BZK15\_N} & \rotatebox{90}{BZK15\_U} & \rotatebox{90}{BZK15} & \rotatebox{90}{BZ18} & \rotatebox{90}{FORD14} \\
                        \midrule
                        MR   & 0.16                                    & 0.08                                   & -0.04                               & 0.01                                                 & -0.06                     & -0.17                    & \textbf{0.39}        & 0.18                 & 0.28                  & 0.09                  & -0.03                    & -0.04                    & -0.05               & 0.05                   & 0.12                     & 0.00                     & -0.02                 & 0.09                 & -0.07                  \\
                        BS   & 0.12                                    & -0.05                                  & -0.05                               & -0.13                                                & -0.11                     & \textbf{-0.33}           & \textbf{0.34}        & 0.07                 & 0.15                  & 0.09                  & -0.22                    & -0.19                    & 0.06                & 0.13                   & 0.04                     & 0.03                     & 0.16                  & 0.08                 & 0.08                   \\
                        RR   & -0.03                                   & -0.01                                  & -0.04                               & -0.23                                                & -0.12                     & -0.29                    & -0.03                & 0.06                 & \textbf{0.43}         & 0.06                  & 0.11                     & 0.13                     & 0.07                & -0.07                  & 0.04                     & 0.10                     & 0.02                  & 0.17                 & -0.09                  \\
                        GSST & 0.18                                    & 0.05                                   & -0.09                               & -0.06                                                & -0.04                     & -0.18                    & \textbf{0.34}        & 0.12                 & \textbf{0.33}         & 0.02                  & 0.16                     & 0.19                     & 0.24                & 0.11                   & 0.08                     & 0.13                     & 0.21                  & 0.14                 & 0.19                   \\
                        GK   & 0.18                                    & 0.01                                   & -0.01                               & -0.13                                                & -0.11                     & -0.12                    & 0.15                 & 0.21                 & \textbf{0.31}         & -0.01                 & -0.02                    & -0.01                    & 0.13                & -0.04                  & 0.12                     & 0.06                     & 0.10                  & 0.17                 & -0.04                  \\
                        NS   & 0.26                                    & 0.04                                   & -0.11                               & -0.09                                                & -0.09                     & -0.17                    & \textbf{0.36}        & -0.19                & 0.27                  & 0.12                  & 0.29                     & \textbf{0.31}            & 0.29                & 0.13                   & 0.07                     & 0.18                     & 0.17                  & 0.19                 & 0.10                   \\
                        JK   & 0.06                                    & -0.02                                  & -0.09                               & -0.10                                                & -0.06                     & -0.26                    & \textbf{0.39}        & 0.09                 & 0.29                  & 0.07                  & 0.13                     & 0.15                     & 0.17                & 0.03                   & 0.05                     & 0.11                     & 0.14                  & 0.18                 & 0.05                   \\
                        BRW  & -0.08                                   & 0.00                                   & -0.04                               & 0.21                                                 & 0.22                      & 0.08                     & \textbf{0.66}        & -0.29                & 0.16                  & \textbf{0.56}         & -0.10                    & -0.10                    & 0.02                & 0.11                   & 0.11                     & -0.08                    & 0.01                  & 0.09                 & -0.22                  \\
                        \bottomrule
                    \end{tabular}%
                    \begin{tablenotes}[para, flushleft]
                        \footnotesize
                        \emph{Note:}   Pairwise correlations between monetary policy proxies (rows) and proxies for other structural shocks (columns). Bold values indicate correlations with an absolute magnitude greater than or equal to 0.30. Column abbreviations denote shock type and source: K21\_S and K21\_N are \cite{kanzig2021macroeconomic} oil supply surprise shocks and oil supply news shocks, respectively. BH19 is the \cite{baumeister2019structural} oil supply shock, defined as the median draw from sign restrictions. OS25, OS25\_out, and OS25\_SR are \cite{ottonello2025financial} financial shocks, corresponding to daily surprises summed over the month, surprises outside the FOMC window, and the sign-restricted structural shock, respectively. RZ18 is the \cite{ramey2018government} military news shock. FP10 is the \cite{fisher2010using} military contractor excess returns. BZP17 is the \cite{benzeev2017chronicle} defense news shock identified using the medium-run horizon method. LRW12 is the \cite{leeper2012quantitative} 1-to-5-year-ahead tax expectation shock. BP06\_SR and BP06\_LR are \cite{beaudry2006stock} TFP news shocks based on short-run and long-run restrictions, respectively, updated by \cite{ramey2016macroeconomic}. F14 is the \cite{fernald2014quarterly} TFP shock, defined as the log difference of TFP, and F14\_U is the utilization-adjusted counterpart. BZK15\_N and BZK15\_U are \cite{benzeev2015investmentspecific} investment-specific technology news shocks and their unanticipated counterparts, respectively. BZK15 is the unanticipated TFP shock in \cite{benzeev2015investmentspecific}. BZ18 is the \cite{benzeev2018what} TFP shock, identified with long-run restrictions. FORD14 is the \cite{francis2014flexible} technology shock, defined as the shock that maximizes the forecast error variance share of labor productivity at a finite horizon $h$.
                    \end{tablenotes}
                \end{threeparttable}}
        \end{table}
    \end{landscape}

    \clearpage
    \newpage
    \section{Proofs}
    We collect some auxiliary lemmas that are useful for the proof of the main theorems below.
    \begin{lemma}[Angular Bounds]\label{lem:angular_bounds}
        The generalized ranking restrictions with proxy quality $\tau\ge0$ in \eqref{eq:rank_O_homog} imply
        \begin{equation}
            \alpha_{\ell,O_{\bullet 1}} \in \bigl[0,\; \rho(\tau)\bigr],\qquad
            \cos(\alpha_{\ell,O_{\bullet 1}})\ge \tau\big| \cos(\alpha_{\ell,O_{\bullet j}})\big|\,,\quad j\ge 2
        \end{equation}
        where $\rho(\tau) \coloneqq \arctan\left(\frac{\sqrt{n-1}}{\tau}\right)\in(0,\frac{\pi}{2})$ with the convention $\rho(\infty)=0$ and $\rho(0)=\frac{\pi}{2}$.
    \end{lemma}

    \subsection{Proofs of Auxiliary Lemmas}
    \begin{proof}[Proof of \autoref{lem:angular_bounds}]
        The second inequality is straightforward, since for all $j\ge 2$,
        \begin{equation}
            \begin{array}{rl}
                \| O_{\bullet 1}\| \| M_{\ell} \| \cos(\alpha_{\ell,O_{\bullet 1}}) & \ge \tau\| O_{\bullet j}\| \| M_{\ell} \| | \cos(\alpha_{\ell,O_{\bullet j}})| \\
                \Longleftrightarrow \cos(\alpha_{\ell,O_{\bullet 1}})               & \ge \tau| \cos(\alpha_{\ell,O_{\bullet j}})|
            \end{array}
            ~.
            \label{eq:rank_angle}
        \end{equation}

        For the first inequality, we first consider the case when $\tau=0$. The generalized ranking restrictions \eqref{eq:rank_O_homog} reduce to
        \begin{equation}
            \|O_{\bullet 1}\|\|M_{\ell}\|\cos(\alpha_{\ell,O_{\bullet 1}})=O_{\bullet 1}'M_{\ell}\ge 0~,
        \end{equation}
        which implies that $\cos(\alpha_{\ell,O_{\bullet 1}})\ge0$ and thus $\alpha_{\ell,O_{\bullet 1}}\in[0,\frac{\pi}{2}]$.

        Next suppose $\tau\in(0,\infty]$. Let $v_{\ell}=O'M_{\ell}$. We have
        \begin{equation}
            \|v_{\ell}\|^2 = v_{\ell}'v_{\ell} = M_{\ell}'OO'M_{\ell}=M_{\ell}'M_{\ell}=\|M_{\ell}\|^2~.
        \end{equation}
        By definition, we have
        \begin{equation}
            \|M_{\ell}\|^2 = v_{\ell,1}^2 + \sum_{j\ge 2}v_{\ell,j}^2\le v_{\ell,1}^2\left[1+\frac{n-1}{\tau^2}\right]
        \end{equation}
        where the inequality follows from the ranking restriction
        \begin{equation}
            O_{\bullet 1}' M_{\ell} = v_{\ell,1}
            \;\ge\;\tau
            \bigl|\,O_{\bullet j}' M_{\ell}\bigr|=\tau|v_{\ell,j}|,
            \quad \text{for all}\; j\ge 2~.
        \end{equation}
        This gives
        \begin{equation}
            \left(1+ \frac{n-1}{\tau^{2}}\right)^{-\frac{1}{2}} \|M_{\ell}\|\le v_{\ell,1}=O_{\bullet 1}'M_{\ell}=\|O_{\bullet 1}\|\|M_{\ell}\|\cos(\alpha_{\ell,O_{\bullet 1}})~,
        \end{equation}
        and thus
        \begin{equation}
            \cos(\alpha_{\ell,O_{\bullet 1}})\ge \left(1+ \frac{n-1}{\tau^{2}}\right)^{-\frac{1}{2}}~.
        \end{equation}
        By definition, $\rho(\tau) = \arctan\left(\frac{\sqrt{n-1}}{\tau}\right)$, and thus
        \begin{equation}
            \begin{array}{rl}
                \tan(\rho)              & =\frac{\sqrt{n-1}}{\tau}                                                                  \\
                \tan^2(\rho)            & =\frac{\sin^2(\rho)}{\cos^2(\rho)}=\frac{1-\cos^2(\rho)}{\cos^2(\rho)}=\frac{n-1}{\tau^2} \\
                (\cos(\rho(\tau)))^{-2} & = 1 + \frac{n-1}{\tau^2}                                                                  \\
                \cos(\rho(\tau))        & =\left(1+ \frac{n-1}{\tau^{2}}\right)^{-\frac{1}{2}}
            \end{array}~,
        \end{equation}
        which gives
        \begin{equation}
            \cos(\alpha_{\ell,O_{\bullet 1}})\ge \cos(\rho(\tau))
        \end{equation}
        and thus the desired result.
    \end{proof}

    \subsection{Proofs of Propositions}

    \begin{proof}[Proof of \autoref{prop:rank_O}]
        First, the proxy contamination assumption~\ref{ass:proxy_tau0} implies the following ranking conditions in correlations:
        \begin{equation}
            \mathrm{corr}(m_{\ell,t},\epsilon_{1,t})
            \;=\;
            \tau_{\ell,0}\,
            \big|\mathrm{corr}(m_{\ell,t},\epsilon_{j,t})\big|,
            \qquad \ell = 1,\ldots,k,\; j\ge 2~,
        \end{equation}
        which, under Assumption \ref{ass:proxy_finite_moments}, is equivalent to
        \begin{equation}
            \mathbb{E}[m_{\ell,t}\epsilon_{1,t}]\ge \tau_{\ell,0} \big|\mathbb{E}[m_{\ell,t}\epsilon_{j,t}]\big|,\quad \text{for all } \ell=1,\ldots,k \text{ and } j\ge 2.\label{eq:grr_mom}
        \end{equation}
        Using $\epsilon_t = O' L^{-1}u_t$ from Assumption \ref{ass:svar}, we have
        \begin{equation}
            \mathbb{E}[m_{\ell,t}\epsilon_{1,t}]
            = \mathbb{E}[e_{1}'\epsilon_{t}m_{\ell,t}]
            =e_{1}'\mathbb{E}[O' L^{-1}u_tm_{\ell,t}]
            =O_{\bullet 1}'\mathbb{E}[L^{-1}u_tm_{\ell,t}]
            =O_{\bullet 1}'M_\ell,
        \end{equation}
        where $O_{\bullet j}$ denotes the $j$-th column of $O$. Similarly, $\mathbb{E}[m_{\ell,t}\epsilon_{j,t}] = O_{\bullet j}' M_\ell$ for $j \geq 2$. Substituting into \eqref{eq:grr_mom} yields \eqref{eq:rank_O}.
    \end{proof}

    \begin{proof}[Proof of \autoref{prop:monotonicity}]
        Since there is a one-to-one mapping from each orthogonal matrix $O$ to the impulse responses $\theta(O)$, the inclusion $\Theta(\tau_{\textsf{large}})\subseteq \Theta(\tau_{\textsf{small}})$ follows directly from $\mathcal{F}(\tau_{\textsf{large}})\subseteq \mathcal{F}(\tau_{\textsf{small}})$. We therefore focus on proving the inclusion for the feasible set of $O$.

        \textbf{Case I:} $\mathcal{F}(\tau_{\textsf{small}})=\varnothing$. Then inclusion holds if we have $\mathcal{F}(\tau_{\textsf{large}})=\varnothing$. We prove this by contradiction.
        Note that under Assumption~\ref{ass:sign}, we have $\mathcal{F}_{\textup{sign}}\neq\varnothing$. Then $\mathcal{F}(\tau_{\textsf{small}})=\varnothing$ must come from violations of the GRR: there does not exist $O\in\mathcal{F}_{\textup{sign}}$ such that
        \begin{equation}
            O_{\bullet 1}'M_\ell\ge\tau_{\textsf{small}}|O_{\bullet j}'M_\ell|,\qquad\text{for all } i\ge 2.
        \end{equation}
        Now suppose $\mathcal{F}(\tau_{\textsf{large}})\neq\varnothing$, then there exists an $\tilde{O}\in\mathcal{F}_{\textup{sign}}$ such that
        \begin{equation}
            \tilde{O}_{\bullet 1}'M_\ell\ge\tau_{\textsf{large}}|\tilde{O}_{\bullet j}'M_\ell|> \tau_{\textsf{small}} |\tilde{O}_{\bullet j}'M_\ell|,\qquad\text{for all } j\ge 2~,
            \label{eq:grr_tau_large_small}
        \end{equation}
        where the first inequality follows from the definition of GRR with $\tau_{\textsf{large}}$ and the second by construction that $\tau_{\textsf{large}}>\tau_{\textsf{small}}$. Given that, we must have $\tilde{O}\in\mathcal{F}(\tau_{\textsf{small}})$, which contradicts the assumption $\mathcal{F}(\tau_{\textsf{small}})=\varnothing$.
        So we must have $\mathcal{F}(\tau_{\textsf{large}})= \mathcal{F}(\tau_{\textsf{small}})=\varnothing$.

        \textbf{Case II:}  $\mathcal{F}(\tau_{\textsf{small}})\neq \varnothing$. If $\mathcal{F}(\tau_{\textsf{large}})=\varnothing$, then $\mathcal{F}(\tau_{\textsf{large}})\subseteq \mathcal{F}(\tau_{\textsf{small}})$ trivially holds. Otherwise, from \eqref{eq:grr_tau_large_small}, any feasible $O\in\mathcal{F}(\tau_{\textsf{large}})$ must also be inside $\mathcal{F}(\tau_{\textsf{small}})$, so we have $\mathcal{F}(\tau_{\textsf{large}})\subseteq \mathcal{F}(\tau_{\textsf{small}})$.

        Combining the two cases gives the desired result.
    \end{proof}

    \subsection{Proofs of Theorems}

    \begin{proof}[Proof of \autoref{thm:tau_identification}]
        Fix any $\tau \in(0,\infty)$ such that the feasible set $\mathcal{F}(\tau)\neq\varnothing$. By definition of feasibility, there exists at least one rotation matrix $\tilde O\in\mathcal{F}(\tau)$. In particular, $\tilde O\in\mathcal{F}_{\textup{sign}}$ and, for every proxy $\ell=1,\ldots,k$, the generalized ranking restrictions hold at $\tilde O$.

        By Lemma~\ref{lem:angular_bounds}, the ranking restrictions imply that for each $\ell$,
        \begin{equation}
            \alpha_{\ell,\tilde O_{\bullet 1}}
            \le
            \rho(\tau)
            \equiv
            \arctan\!\left(\frac{\sqrt{n-1}}{\tau}\right).
        \end{equation}
        Taking the maximum over $\ell$ yields
        \begin{equation}
            \max_{\ell=1,\ldots,k}
            \alpha_{\ell,\tilde O_{\bullet 1}}
            \le
            \rho(\tau).
        \end{equation}

        Consider the quantity $\min_{q\in\mathcal{G}_{\textup{sign}}}\max_{\ell=1,\ldots,k}\alpha_{\ell,q}$, which measures the worst-case misalignment between any sign-feasible target column $q\in\mathcal{G}_{\textup{sign}}$ and the collection of proxy covariance vectors $\{M_{\ell}\}_{\ell=1}^{k}$. We then have
        \begin{equation}
            \min_{q\in\mathcal{G}_{\textup{sign}}}
            \max_{\ell}
            \alpha_{\ell,q}
            \le
            \max_{\ell}
            \alpha_{\ell,\tilde O_{\bullet 1}}
            \le
            \rho(\tau).
        \end{equation}
        The first inequality is obtained by construction due to the $\min$ operator and that $\tilde{O}\in\mathcal{G}_{\textup{sign}}$; the second comes from Lemma~\ref{lem:angular_bounds}: for any $\ell$, the feasible $\tilde{O}_{\bullet 1}$ must be upper bounded by $\rho(\tau)$.

        Because the cotangent function is strictly decreasing on $(0,\pi/2]$, it follows that
        \begin{equation}
            \cot\!\left(
            \min_{q\in\mathcal{G}_{\textup{sign}}}
            \max_{\ell}
            \alpha_{\ell,q}
            \right)
            \ge
            \cot(\rho(\tau)).
        \end{equation}
        Using the definition of $\rho(\tau)$, we have $\cot(\rho(\tau))=\tau/\sqrt{n-1}$, and hence
        \begin{equation}
            \tau
            \le
            \sqrt{n-1}\,
            \cot\!\left(
            \min_{q\in\mathcal{G}_{\textup{sign}}}
            \max_{\ell}
            \alpha_{\ell,q}
            \right).\label{eq:tau_bound_in_proof}
        \end{equation}

        It remains to show that \eqref{eq:tau_bound_in_proof} holds when $\tau=0$ and $\tau=\infty$.
        First, when $\tau=0$ the above holds trivially since $\alpha_{\ell,q}\in[0,\frac{\pi}{2}]$ and thus
        \begin{equation*}
            \sqrt{n-1}\,
            \cot\!\left(
            \min_{q\in\mathcal{G}_{\textup{sign}}}
            \max_{\ell}
            \alpha_{\ell,q}
            \right)\ge \sqrt{n-1}\,
            \cot\!\left(\frac{\pi}{2}
            \right)=0=\tau~.
        \end{equation*}
        Second, when $\tau=\infty$, Lemma~\ref{lem:angular_bounds} shows that we must have $\alpha_{\ell,q}=0$---the target column of interest is point-identified. We have
        \begin{equation*}
            \sqrt{n-1}\,
            \cot\!\left(
            \min_{q\in\mathcal{G}_{\textup{sign}}}
            \max_{\ell}
            \alpha_{\ell,q}
            \right)=\sqrt{n-1}\,
            \cot\!\left(
            0
            \right)=\infty=\tau~.
        \end{equation*}

        Finally, we need to show that \eqref{eq:tau_bound_in_proof} holds for the true $\tau_0$.
        Assumption~\ref{ass:proxy_tau0} implies that the true proxy quality $\tau_0$ must itself be feasible, i.e., $\mathcal{F}(\tau_0)\neq\varnothing$. Applying the above argument at $\tau=\tau_0$ yields $\tau_0\le\bar\tau$, completing the proof.
    \end{proof}

    \begin{proof}[Proof of \autoref{thm:point_identification}]
        The proof is constructive: for any given $ \tau_{0}\ge 1 $, we will construct a set of $ k=2 $ proxy vectors $ M_1 $ and $ M_2 $ that uniquely pins down the first column of the rotation matrix $ O_{\bullet 1} $. Then, we augment the set of proxies by adding more vectors that do not change the feasible set.

        By \autoref{lem:angular_bounds}, we know that the generalized ranking restrictions for a single proxy $ M_{1} $ confines $ O_{\bullet 1} $ to a spherical cap centered at $ M_{1} $, with the angle between $ O_{\bullet 1} $  and $ M_{1} $ upper bounded by $\rho(\tau_0)$, which gives
        \begin{equation}
            \cos \rho(\tau_0) = \sqrt{\frac{\tau_0^{2}}{\tau_0^{2}+n-1}},\quad
            \sin \rho(\tau_0) = \sqrt{\frac{n-1}{\tau_0^{2}+n-1}}~.
        \end{equation}
        We can then construct proxies $ M_1 $ and $ M_2 $ by
        \begin{equation}
            \begin{array}{rl}
                M_1 = & (\cos\rho(\tau_0))O_{0,\bullet 1}  + \frac{\sin \rho(\tau_0)}{\sqrt{n-1}}\sum_{i\ge 2} O_{0,\bullet j} \\
                M_2 = & (\cos\rho(\tau_0))O_{0,\bullet 1}  - \frac{\sin \rho(\tau_0)}{\sqrt{n-1}}\sum_{i\ge 2} O_{0,\bullet j}
            \end{array}~.
        \end{equation}
        It remains to verify that the optimization problem \eqref{eq:optim_obj} with constraints \eqref{eq:constraints} has a unique solution $ O_{0} $ when the two proxies $ M_1 $ and $ M_2 $ are used.

        First, we verify that $ O_{0} $ is feasible. By construction, $ O_{0} $ satisfies the sign restrictions. Moreover, we have
        \begin{equation*}
            O_{0,\bullet 1}'M_1 = \cos\rho(\tau_0) \ge \tau_{0}\frac{\sin\rho(\tau_0)}{\sqrt{n-1}} = \tau_{0}|O_{0,\bullet j}'M_1|,\quad \text{for all } i\ge 2~,
        \end{equation*}
        and
        \begin{equation*}
            O_{0,\bullet 1}'M_2 = \cos\rho(\tau_0) \ge \tau_{0}\frac{\sin\rho(\tau_0)}{\sqrt{n-1}} = \tau_{0}|O_{0,\bullet j}'M_2|,\quad \text{for all } i\ge 2~.
        \end{equation*}
        Therefore, $ O_0 $ satisfies the generalized ranking restrictions for both proxies. Taken together, we have $ O_0\in \mathcal{F}(\tau_0) $.

        Second, we show that $ O_{0} $ is the unique feasible solution. Suppose there exists another feasible solution $ O\in \mathcal{F}(\tau_0) $. Then by \autoref{lem:angular_bounds}, we have
        \begin{equation*}
            \alpha_{1,O_{\bullet 1}}\le \rho(\tau_0),\quad
            \alpha_{2,O_{\bullet 1}}\le \rho(\tau_0)~.
        \end{equation*}
        By the triangle inequality, we have
        \begin{equation*}
            2\rho(\tau_0)=\alpha_{1,M_2} \le \alpha_{1,O_{\bullet 1}} + \alpha_{2,O_{\bullet 1}} \le 2\rho(\tau_0)
        \end{equation*}
        where the first equality follows from the construction of $ M_1 $ and $ M_2 $:
        \begin{equation*}
            M_1'M_{2} = \cos^{2}\rho(\tau_0) - \sin^{2}\rho(\tau_0) = \cos(2\rho(\tau_0))~.
        \end{equation*}
        As a result, we must have $ \alpha_{1,O_{\bullet 1}}=\alpha_{2,O_{\bullet 1}}=\rho(\tau_0) $ for any feasible $ O_{\bullet 1} $. Given that, triangle inequality holds with equality, which implies that $ O_{\bullet 1} $ must lie on the plane spanned by $ M_1 $ and $ M_2 $, i.e.,
        \begin{equation*}
            O_{\bullet 1} = a M_1 + b M_2~.
        \end{equation*}
        Combining the two angle conditions, we have
        \begin{equation*}
            \begin{array}{rl}
                \cos\rho(\tau_0) = & O_{\bullet 1}'M_1 = a + b\cos(2\rho(\tau_0))   \\
                \cos\rho(\tau_0) = & O_{\bullet 1}'M_2 = a\cos(2\rho(\tau_0)) + b~.
            \end{array}
        \end{equation*}
        Solving the above equations gives
        \begin{equation*}
            a = b = \frac{1}{2\cos\rho(\tau_0)}~.
        \end{equation*}
        Therefore, we have
        \begin{equation*}
            O_{\bullet 1} = \frac{1}{2\cos\rho(\tau_0)}(M_1 + M_2) = O_{0,\bullet 1}~.
        \end{equation*}
        Thus, $ O_{0,\bullet 1}$ is the unique feasible solution for the $j$-th column of $ O $, when the two proxies $ M_1 $ and $ M_2 $ are used.

        Finally, we can keep adding more proxies $ M_{3},\dots, M_{k} $ such that the corresponding spherical caps contain $ O_{0,\bullet 1} $, i.e., $ \alpha_{\ell, O_{0,\bullet 1}} \le \rho(\tau_0) $ for all $ \ell\ge 3 $. This will not change the feasible set.
    \end{proof}

\end{appendices} 

\end{document}